\documentclass[9pt]{elife}

\usepackage{lipsum} 
\usepackage[version=4]{mhchem}
\usepackage{siunitx}
\DeclareSIUnit\Molar{M}

\usepackage{ifthen}
\newcommand{\TODO}[1]{%
    {\color{magenta}\textbf{TODO}%
        \ifthenelse{\equal{#1}{}}{}{: #1}%
    }
}

\usepackage[many, listings]{tcolorbox}

\NewDocumentCommand{\PLACEHOLDER}{m}{{
    \
    \begin{tcolorbox}[
        colframe=black!40,
        colback=black!5,
    ]%
        \normalfont\textit{Placeholder%
        {\ifthenelse{\equal{#1}{}}{}{ $-$ #1}}%
    }
    \end{tcolorbox}
}}

\newtcolorbox{note}[1][]{
    title=NOTE,
    float=ht,
    fonttitle=\bfseries,
    coltitle=black,
    colframe=teal!40,
    colback=teal!5,
    before skip=1em plus 2pt,
    after skip=1em plus 2pt,
    #1
}

\tcbset{before upper=\setlength{\parskip}{1em}}

\definecolor{contentboxcolor}{RGB}{20,105,176}

\newtcolorbox[
    auto counter
]{contentbox}[2][]{
    title=Box~\thetcbcounter. {\normalfont\itshape #2},
    fonttitle=\bfseries,
    breakable,
    enhanced,
    before skip=1em plus 5pt,
    after skip=1em plus 5pt,
    coltitle=white,
    colframe=contentboxcolor,
    colback=contentboxcolor!5,
    #1
}

\usepackage{adjustbox} 
\setlist{leftmargin=1.5em, nosep} 
\usepackage{tikz} 
\usetikzlibrary{trees}
\usepackage{adjustbox} 
\usepackage{fontawesome} 
\usepackage{dirtytalk} 
\usepackage{pgfgantt} 
\usepackage[normalem]{ulem} 

\usepackage{xurl}

\usepackage{soul}
\colorlet{background}{black!20!white!80!}
\sethlcolor{background}

\newcommand{\keypoints}[1]{}

\definecolor{neural}{HTML}{1B9E77}
\definecolor{behavior}{HTML}{D95F02}
\definecolor{ext}{HTML}{7570B3}
\tikzstyle{neural}=[draw=neural]
\tikzstyle{behavior}=[draw=behavior]
\tikzstyle{ext}=[draw=ext]

\usepackage{placeins} 

\usepackage{comment}

\renewcommand{\refname}{References}
\makeatletter
\renewcommand{\bibsection}{%
   \section*{\refname
            \@mkboth{{\refname}}{{\refname}}%
   }%
   \addcontentsline{toc}{section}{\refname} 
}
\makeatother

\usepackage{listings}  
\definecolor{mygreen}{rgb}{0,0.6,0}
\definecolor{mygray}{rgb}{0.5,0.5,0.5}
\definecolor{mymauve}{rgb}{0.58,0,0.82}
\usepackage{parskip} 

\usepackage{subcaption}  
\usepackage[abbreviations, UKenglish]{foreign}  

\addto\extrasenglish{%
}

\usepackage{datetime2}  

\title{A perspective on neuroscience data standardization with Neurodata Without Borders}

\author[12\authfn{1}]{Andrea Pierré}
\author[12\authfn{1}]{Tuan Pham}
\author[3]{Jonah Pearl}
\author[3]{Sandeep Robert Datta}
\author[2\authfn{1}*]{Jason T. Ritt}
\author[12*]{Alexander Fleischmann}

\affil[1]{Department of Neuroscience, Division of Biology and Medicine, Brown University, Providence, USA}
\affil[2]{The Robert J. and Nancy D. Carney Institute for Brain Science, Brown University, Providence, USA}
\affil[3]{Department of Neurobiology, Harvard Medical School, Boston, USA}

\corr{jason_ritt@brown.edu}{JTR}
\corr{alexander_fleischmann@brown.edu}{AF}

\contrib[\authfn{1}]{These authors contributed equally to this work}

\begin{document}

\maketitle

\begin{abstract}
  Neuroscience research has evolved to generate increasingly large and complex experimental data sets, and advanced data science tools are taking on central roles in neuroscience research. Neurodata Without Borders (NWB), a standard language for neurophysiology data, has recently emerged as a powerful solution for data management, analysis, and sharing. We here discuss our labs’ efforts to implement NWB data science pipelines. We describe general principles and specific use cases that illustrate successes, challenges, and non-trivial decisions in software engineering. We hope that our experience can provide guidance for the neuroscience community and help bridge the gap between experimental neuroscience and data science.
\end{abstract}

\tableofcontents

\section{Introduction}

\subsection{Increasing complexity of neuroscience data}%
\label{sec:big-data-neuroscience}

Over the past 20 years, neuroscience research has been radically changed by two major trends in data production and analysis. First, neuroscience research now routinely generates large datasets of high complexity. Examples include recordings of activity across large populations of neurons, often with high resolution behavioral tracking~\citep{steinmetz_distributed_2019, stringer_spontaneous_2019, mathis_deeplabcut_2018, siegle_survey_2021, koch_next-generation_2022}, analyses of neural connectivity at high spatial resolution and across large brain areas~\citep{scheffer_connectome_2020, loomba_connectomic_2022}, and detailed molecular profiling of neural cells~\citep{yao_high-resolution_2023, langlieb_cell_2023, braun_comprehensive_2022, callaway_multimodal_2021}. Such large, multi-modal data sets are essential for solving major questions about brain function~\citep{brose_global_2016, jorgenson_brain_2015, koch_big_2016}.

Second, the collection and analysis of such datasets requires interdisciplinary teams, incorporating expertise in systems neuroscience, engineering, molecular biology, data science, and theory. These two trends are reflected in the increasing numbers of authors on scientific publications~\citep{wareham_new_2016}, and the creation of mechanisms to support team science by the NIH and similar research funding bodies~\citep{cooke_enhancing_2015, volkow_enhancing_2022, brose_global_2016}.

There is also an increasing scope of research questions that can be addressed by aggregating ``open data'' from multiple studies across independent labs. Funding agencies and publishers have begun to aggressively promote data sharing and open data, with the goals of improving reproducibility and increasing data reuse~\citep{dallmeier-tiessen_enabling_2014, tenopir_changes_2015, pasquetto_reuse_2017}.
However, open data may be unusable if scattered in a wide variety of naming conventions and file formats lacking machine-readable metadata. 

Big data and team science necessitate new strategies for how to best organize data, with a key technical challenge being the development of standardized file formats for storing, sharing, and querying datasets. Prominent examples include the Brain Imaging Data Structure (BIDS) for neuroimaging, and Neurodata Without Borders (NWB) for neurophysiology data~\citep{teeters_neurodata_2015, gorgolewski_brain_2016, rubel_neurodata_2022, holdgraf_ieeg-bids_2019}. The Open Neurophysiology Environment (ONE), best known from adoption by The International Brain Laboratory~\citep{the_international_brain_laboratory_data_2020, the_international_brain_laboratory_modular_2023}, has a similar application domain to NWB, but a highly different technical design. These initiatives provide technical tools for storing and accessing data in known formats, but more importantly provide conceptual frameworks with which to standardize data organization and description in an (ideally) universal, interoperable, and machine-readable way.

\subsection{Our labs’ history in implementing NWB-based standardization}

In 2019, the Fleischmann and Ritt labs initiated a collaboration to enhance the Fleischmann lab's data science and computational tooling and workflows. We expanded our team by hiring two research software engineers (RSE), and by extending collaborations with data scientists and computational biologists. Similar efforts were underway in the Datta lab. An early common goal was the standardization of neurophysiology and behavioral data using a framework such as NWB. In this manuscript, we provide our perspective on opportunities and challenges when adopting NWB data standardization.

\begin{figure}
\begin{fullwidth}
  \centering
  \includegraphics[width=0.9\linewidth]{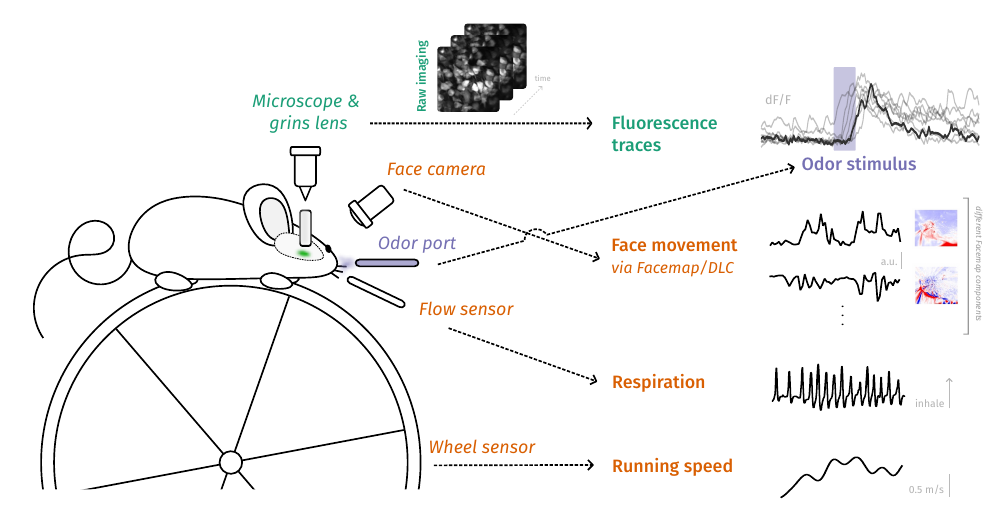}
  \caption[Setup of a typical experiment and resulting data streams]{%
      \textbf{Setup of a typical Fleischmann lab experiment and resulting data streams.}
      The left schematic illustrates \textit{in vivo} head-fixed two-photon calcium imaging of a deep brain area (\eg{} piriform cortex) through a GRIN lens.
      Throughout the paper, we use the following color scheme: green for neural activity, orange for animal behaviors, and purple for external variables (\eg{} stimuli).
      Raw images from the microscope (top) are preprocessed to obtain fluorescence time series for each segmented neuron (top row, right).
      The animal receives odor stimuli through an odor port during a time window in each trial, marked by a light purple bar in the fluorescence time series plot.
      Several behaviors are tracked.
      A high resolution camera captures facial movement, typically reduced using the application Facemap into principal components of image motion (middle), or through DeepLabCut into pose estimation or keypoints.
      Peri-nasal flow and wheel sensors, connected through a microcontroller, provide respiration and running speed estimates, respectively.%
  }%
  \label{fig:exp-schematic}
  \vspace{1em}
\end{fullwidth}
\end{figure}

Our labs investigate the functions of neural circuits for sensory processing and behavior in mice. Typical experiments include calcium imaging of neuronal activity in awake, head-fixed mice during odor presentation, with a number of behavioral readouts including sniffing, running, and facial movements (see \autoref{fig:exp-schematic}). In other experiments, mice are freely moving, with implanted GRIN lenses for miniscope imaging, odor and reward delivery in nose ports, and behavioral readouts including videographic tracking. Our experimental designs, data generation, and analyses are similar to many other labs investigating neural circuit mechanisms for sensory-motor transformations, learning, and memory (\autoref{box:lab_workflow}), though each lab has its own idiosyncrasies impinging on data management.

In this manuscript, we first discuss our motivation and general considerations for implementing data standardization. We then describe the implementation of NWB data conversion pipelines, including domain-specific use cases and solutions for data sharing. We conclude by identifying opportunities for improving future user experience. We hope that by describing our experience, other labs planning to adopt NWB will benefit from comparisons with their own needs and capabilities. We also hope to provide a case study that may be informative for developers of NWB and similar data science toolboxes.

\begin{featurebox}
\caption{Fleischmann Lab workflow}%
\label{box:lab_workflow}

    \textbf{Data Acquisition --- Experiments and Systems}:
        We perform \textit{in vivo} calcium imaging experiments in head-fixed (2-photon imaging) and freely moving (miniscope) mice. Experiments include multi-plane, multi-color, and/or multi-day recordings.

    \textbf{Data Acquisition --- Tasks and Stimuli}:
        In some experiments, animals receive pre-programmed odor stimuli independent of their behavior; in other experiments, sensory stimuli or an animal's behavior can trigger a reward. Behavior recording includes micro-controller-acquired time series (\eg{} wheel speed, sniff rate, licks, rewards) and video recordings of the animal's face or body motion.

        \textbf{Preprocessing}: Pipelines include conventional calcium imaging steps (\eg{} motion correction, segmentation, deconvolution, multi-color or multi-day registration) using existing tools such as Suite2p~\citep{pachitariu_suite2p_2016} and Inscopix\footnote{\url{https://www.inscopix.com/}}
        Experiments with behavioral videos may also be preprocessed with toolboxes such as DeepLabCut~\citep{mathis_deeplabcut_2018} for pose estimation and Facemap~\citep{syeda_facemap_2022} for facial motion extraction.

    \textbf{Conversion to standard format}: Raw and preprocessed data streams are integrated and stored in NWB files, using a custom tool, \texttt{calimag}~\citep{pierre_calimag_2023}, developed in the Fleischmann lab.

    \textbf{Analyses}: Questions include stimulus or behavior tuning of single neuron or population activity, as well as how learning and experience shape neural activity.
\end{featurebox}

\section{Key stakeholders in adoption of a new lab standard}
\label{sec:key_stakeholders}

We first define, in high level terms, three distinct personnel roles in a typical research lab, each of whom has their own needs and incentives surrounding data standardization:
\begin{itemize}
    \item \textbf{PIs} are
        principal investigators and senior researchers that manage research teams, labs, and projects.
    \item \textbf{Researchers} include
        research trainees (\eg{} undergraduate and graduate students, postdoctoral associates), lab technicians, and data scientists, and more generally individuals collecting and/or analyzing data.
    \item \textbf{Research software engineers} (RSE) support researchers by developing and maintaining software, packages, and pipelines for data management, processing, and analysis.
\end{itemize}

\subsection{PIs}
Key desired outcomes for the adoption of lab-wide standardized data formats include improved efficiency, rigor, reproducibility, and ease of collaboration. Efficiency could follow from using common tools for saving, retrieving, analyzing, and sharing data; technical improvements by one member can have knock-on value for others. Rigor and reproducibility similarly benefit from increased access and scrutiny brought by all lab members being able to see each other’s work, instead of working in isolation; data already in standard formats could ease communication and usage. An additional value for PIs is meeting the norms of their field for data management and sharing, including mandates from funding agencies such as the NIH, without requiring extensive \textit{ad hoc} effort at the time of grant submissions or publication.
\keypoints{key objectives: efficiency, rigor, reproducibility, collaboration}

However, there are several concerns when introducing standardized formats. PIs generally want to avoid major disruptions to scientific productivity in the lab. There is rarely a good time to slow or halt data collection and analysis in order to fully convert to new pipelines and workflows. On the other hand, a gradual transition can paradoxically lead to greater friction due to the simultaneous use of multiple incompatible systems. Adoption of a data standard can be much more than a point-and-click operation, requiring many decisions about the structure and use of the data not just as it is now, but also what the PI expects it to be in the future. One of the first decisions is the standard itself: it can be difficult to pick a “winner”, as standards may quickly become incompatible with the lab’s evolving methods.
\keypoints{challenge: disrupt productivity}

It is also uncommon to have institutional support, in the form of grant funding or university staffing allocated to the “low level” task of revising data formats, or incentives such as promotion criteria that reward best practices in data management. While research software engineers (RSE) are increasingly recognized as valuable contributors to the research enterprise \citep{carver_survey_2022}, most labs still do not have access to an RSE. This places the burden on students and postdocs, who are often enthusiastic to adopt new practices but are constrained by a need to make continual progress in their own careers. Moreover, lab members, including PIs, generally lack advanced training to know how to build automated systems that integrate multiple data streams into a single format with appropriate metadata, provide that data for analysis, and share data following community norms such as FAIR guidelines \citep{wilkinson_fair_2016}. Without support, adopting a standard is often a shared aspiration with little personal buy-in to do the needed work.
\keypoints{challenge: support, funding}

\subsection{Researchers}
\label{sec:researcher-perspectives}

The main motivation for researchers to adopt standardized data formats is to improve data analysis and shareability. Standardized data formats may support efficient and reproducible data processing and flexible, comprehensive data exploration and analysis. Efficient data analysis can, in turn, provide critical information for optimizing experimental design. Furthermore, standardized formats facilitate data sharing, which can yield new perspectives on datasets and increase their impact.
\keypoints{main goal: more efficient analysis}

A main concern is that data standardization requires a significant increase in workload, whether researchers tackle it on their own or in collaboration with an RSE. The increased workload can happen at the experiment and data conversion stages, if data management standardization comes at the expense of experimental flexibility. At the stage of analysis, researchers may need to spend time to learn and adapt to the new standard in order to use the data. Researchers' diverse backgrounds, the availability/support of tools for standardized data, and the maturity of their projects further contribute to tradeoffs between making consistent experimental progress and standardizing experimental outputs. In particular, there are limited training opportunities in scientific computing as a topic in its own right, leaving most researchers without conceptual frameworks and technical knowledge to properly guide these choices. Additionally, researchers who decide to embrace standardization, open data, and reproducible workflows often lack recognition for the added work.
\keypoints{standardization $==$ increased workload}
\keypoints{lack of incentives}

\subsection{Research software engineers}
RSEs directly support researchers in data management, analysis, sharing, and publication. Adopting standardized formats establishes predictability in the data that the researchers produce. This facilitates communication and makes it easier for RSEs to efficiently provide support in finding, using, and building appropriate systems to interact with the data. RSEs can also take advantage of such predictability to provide sufficient documentation and usable examples of the data for analysis, sharing and re-use.
\keypoints{benefits: ease in support}

A core challenge is developing stable software implementations and workflows that are robust to small variations in experimental data, while still allowing flexibility to be useful to researchers engaged in rapid evolution of diverse experimental designs. Furthermore, choosing a new technology carries an elevated risk of bugs and missing features. Open source tools can be particularly unpredictable, and extensive in-house workarounds may be unsustainable and defeat the original purpose of standardization.
\keypoints{challenge: evolvability}
\keypoints{unpredictability of open source \& young tech}

In addition, researchers and RSEs often come from different backgrounds. RSEs may not be familiar with scientific priorities and experimental constraints, and the expectations and timeline of research projects. Thus, diverging expectations and miscommunication between researchers and RSEs can lead to friction and delay in adopting the standards.
\keypoints{challenge: different expectations}

\section{Social scales of working with the NWB standard}%
\label{sec:social_scales}

\subsection{Within a lab}

It is often desirable for members of a lab to share and use common technology, including analysis code, data conversion pipelines, and/or acquisition systems. This commonality allows members to jointly address technical problems, and build on top of known solutions with some degree of prior validation, creating consistency across \say{generations} of graduate students and postdocs. For example, in our lab, researchers performing head fixed two-photon calcium imaging share the same acquisition systems and data conversion pipeline, which allows them to get advice from their peers and to contribute their own solutions to common pain-points.

A potential pitfall of sharing a common set of technologies may arise when the technology is not well maintained or kept up-to-date, forcing new projects to build on shaky ground. Another pitfall may come from the complexity of supporting a diverse enough set of use cases, and trying to make them all fit into the same technology.
\keypoints{sharing of tech across lab members}

On-boarding is key to encourage this economy of scale and self-regeneration of benefits, especially if a standard is not yet established. For example, rather than introduce NWB to researchers in new analysis notebooks, we tried to work backwards from the analysis pipelines they already used. That is, we refactored researchers' existing code by replacing only file load operations, and converting from NWB structure to whatever variable names and data types the researchers already used (that often adopted suboptimal data conventions from the original raw file formats). Further experience with NWB might motivate changes to those conventions, but in this approach, initial learning is focused on practical steps whose value is innately recognized by the researcher, rather than on the generic NWB software interface. Naturally, it could be simpler for new lab members (or new projects) to start from a standardized ``clean slate'', though our experience is that in practice there is usually still substantial inheritance of older code and procedures, at least in an established lab.

The Fleischmann lab uses lab-wide Git hosting (on GitLab), facilitating internal sharing and collaborative development of code. Combined with regular lab meeting discussion of data management and analysis topics, this culture of open communication and sharing helps disseminate technical progress across all lab members.

\subsection{Collaboration}%
\label{sec:researcher:collab}

Our experience using NWB to send data to collaborators in other labs has been more mixed than for internal adoption. While standardization aims to establish a universal language for data, there can still be friction for recipients who have not already installed and used the necessary software, especially in the absence of good documentation and relevant working examples. We describe two cases with two different labs performing additional analyses on data we collected.
\keypoints{intro to challenge: maturity of project + collaborator's experience}

In the first case, we provided our collaborators with raw microscope images as TIFF stacks and pre-processed calcium activity time series in NWB format. In contrast to our naive initial expectations, it was challenging for our collaborators to learn how to work with the NWB files. With hindsight, we should have included working example code that loaded and displayed data, which they could use as a starting template for their own work. However, there would still have been some friction, as their lab works primarily in Matlab, while we work almost entirely in Python. NWB provides APIs for both environments, but we would have needed to generate example code from scratch, and the two labs would have maintained two separate code bases. In the end our collaborators used only the TIFF stacks, though partly in order to also work on novel pre-processing algorithms.

In the second case, our collaborators had previous experience with NWB. However, we were still refining our NWB conversion of that data, and were regularly making code breaking changes. Hence, we chose to create and send python \say{pickle} files that contained only a subset of the data, organized to simplify usage on their end and make it easier for us to create example code and documentation. As we continued to develop our internal pipelines, this approach hampered code interoperability between our labs. However, it was the more expedient choice to get the collaborators up and running. We are working to improve the long term stability of our NWB conversion pipeline, in order to converge on a collaboration strategy built entirely on NWB standardization.

\subsection{Public data sharing}

Researchers are increasingly asked to publish their data on public archives. Apart from publication and funding requirements and opportunities for collaboration, these public data repositories increase chances of data reuse, \eg{} for education, benchmarking new tools, computational modelling, or meta-analysis. Popular repositories include Figshare\footnote{\url{https://figshare.com/}},
Zenodo\footnote{\url{https://zenodo.org/}},
OSF~\citep{foster_open_2017} and GIN G-Node\footnote{\url{https://gin.g-node.org/}}.
These are more general repositories, with limited restrictions on data formats, though there sometimes could be other logistical/funding complications.

The Distributed Archives for Neurophysiology Data Integration (DANDI~\citep{halchenko_dandidandi-cli_2022}) is the recommended choice for public data sharing of NWB datasets, and is supported by both the BRAIN Initiative~\citep{kaiser_nihs_2022} and the AWS Public dataset programs.
While it is more restrictive compared to other repositories (for example DANDI allows only standardized formats\footnote{\url{https://www.dandiarchive.org/handbook/about/policies/}},
while Zenodo allows all formats\footnote{\url{https://about.zenodo.org/policies/}}),
the resulting rigor and consistency from DANDI may better facilitate reproducibility, modelling, meta-analysis, and tool development~\citep{dichter_dandi_2023, dichter_neurosift_2023}. We discuss our experience contributing a demonstrative calcium imaging dataset~\citep{daste_two_2022} on DANDI in \nameref{sec:dandi-surprises}.

Apart from file format restrictions, researchers may need to take into account file size limits. DANDI has fairly generous limits, with 5 TB per file and no limit on dataset size, while some repositories have limits of less than 100 GB per file or dataset (some offer higher limits for a fee or other arrangement).

\subsection{NWB community}

During the process of developing our NWB data conversion pipeline, we had several opportunities to interact with the NWB development team. Some of these ways were the NWB/DANDI Slack for quick questions, GitHub issues for a technical question or bug, GitHub discussions for entry level questions, remote meetings with the NWB team for more in-depth substantial guidance, and organized events (hackathons, user days, data re-hack) to meet others from the community and learn about the progress of the ecosystem. In general, our interactions with the NWB community were friendly, helpful, and responsive. For example, our questions on Slack usually received responses within the day. From our observation, this was also true for questions posed by other users.

As described in \nameref{sec-NWB-extensions}, we decided to design our own NWB extensions, which was technically challenging. Communication and assistance from the NWB team was very valuable in our design and implementation. Occasionally there were also helpful examples in GitHub issues or discussions on GitHub and Slack.

That said, many of these resources and communication channels are more familiar to computational scientists and software developers. The official documentation sometimes could be overwhelming to navigate (see, \eg{}~\citep{saunders_documentation_2022}), increasing a typical user's need to find and access these discussions scattered around many channels. It could have been helpful to have a centralized, searchable resource that aggregated and archived these different issues and discussions across different forums, as a complement to the official documentation.

\subsection{Neuroscience community}

The advent of the open science movement, in parallel with standards development, has increased access to software tools and data that until recently was generally limited to high resource institutions. For example, the Allen Institute for Brain Science released an SDK that simplifies retrieval of and interaction with extensive collections of NWB standardized data recorded with cutting edge electrophysiology and imaging tools. Such initiatives greatly expand opportunities to reuse data in education\footnote{\url{https://training.incf.org/collection/neurodata-without-borders-neurophysiology-nwbn}}~\citep{voytek_teaching_2020, van_viegen_neuromatch_2021},
basic research~\citep{deitch_representational_2021}, and bench-marking of new computational models~\citep{schneider_transcriptomic_2023}.

\begin{figure}
\begin{fullwidth}
    \centering
    \begin{tikzpicture}[%
        grow via three points={one child at (0.5,-0.7) and
        two children at (0.5,-0.7) and (0.5,-1.4)},
        edge from parent path={(\tikzparentnode.south) |- (\tikzchildnode.west)}]
    \tikzstyle{every node}=[draw, thick, anchor=west, rounded corners=0.1cm]

    \node [draw=none] {\includegraphics[trim=0.5cm 11cm 0cm 1cm, width=\linewidth]{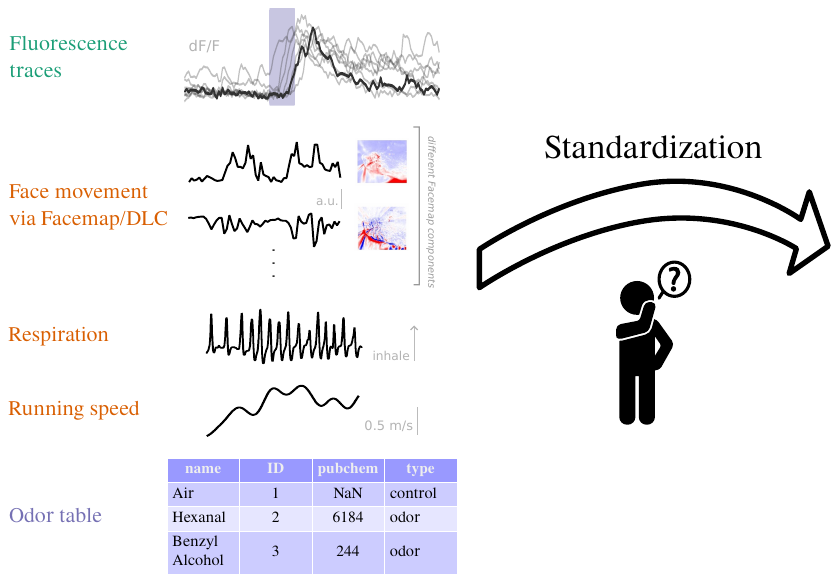}};

    \begin{scope}[scale=0.7, transform shape]
        \node [draw=none] (title) at (12,4) {\textbf{Data lineage focused}};
        \node [draw=none] (title2) at (12,3.7) {NWB solution};
        \node at (12, 3) {\faDatabase~Data file}
            child {node {\faFolderOpen~acquisition}
                child {node [behavior] {\faAreaChart~Running}}
                child {node [behavior] {\faAreaChart~Sniffing}}
                child {node [ext] {\faTable~Odor table}}
                child {node [neural] {\faImage~ImageSeries}}
            }
            child [missing] {}
            child [missing] {}
            child [missing] {}
            child [missing] {}
            child {node {\faFolderOpen~processing}
                child {node [neural] {\faFolderOpen~ophys}
                    child {node [neural] {\faAreaChart~Fluorescence}}
                }
                child [missing] {}
                child {node [behavior] {\faFolderOpen~behavior}
                    child {node [behavior] {\faAreaChart~Poses}}
                    child {node [behavior] {\faAreaChart~Face components}}
                }
                }
        ;
    \end{scope}

    \begin{scope}[scale=0.7, transform shape]
        \node [draw=none] (title) at (19,4) {\textbf{Category focused}};
        \node [draw=none] (title2) at (19,3.7) {Potential alternative};
        \node at (19, 3) {\faDatabase~Data file}
            child {node [neural] {\faFolderOpen~neural}
                child {node [neural] {\faImage~ImageSeries}}
                child {node [neural] {\faAreaChart~Fluorescence}}
            }
            child [missing] {}
            child [missing] {}
            child {node [behavior] {\faFolderOpen~behavior}
                child {node [behavior] {\faAreaChart~Running}}
                child {node [behavior] {\faAreaChart~Sniffing}}
                child {node [behavior] {\faAreaChart~Poses}}
                child {node [behavior] {\faAreaChart~Face components}}
                }
            child [missing] {}
            child [missing] {}
            child [missing] {}
            child [missing] {}
            child {node [ext] {\faFolderOpen~external}
                child {node [ext] {\faTable~Odor table}}
            }
        ;
    \end{scope}
   \end{tikzpicture}
\caption[The issue of data standardization]{\textbf{The issue of data standardization.} Systems neuroscience data tends to be multi-modal, \eg{} time series recorded from standalone sensors and extracted from neural imaging and behavioral videos, plus tables of stimulus or other events (left column). These data are usually scattered across different files in various formats. Researchers wanting a unified standard for ease of analysis and data sharing must choose between at least two possible organizational strategies: prioritizing the data lineage (chosen by NWB format; middle column) or prioritizing  conceptual categories of data sources (right column). Color scheme: green for neural activity, orange for animal behavior, and purple for external variables (\eg{} stimulus).}
\vspace{1.5em}%
\label{fig:data-standard}
\end{fullwidth}
\end{figure}


However, given differences in cultures, priorities, resources, and incentives across different labs and institutions, adoption of NWB, and of open science practices more generally, remains challenging. Institutional policies like the recently updated NIH Data Management Policy~\citep{nih_office_of_intramural_research_2023_2023} add new expectations for researchers, but without creating meaningful recognition and training to support and encourage changes in their practice. Individual institutions also have historically provided minimal support for adoption of data management best practices. We advocate for better funding for standardization as an essential practice in science in general, and particularly for NWB adoption. Some of this support could include partnerships with public resources such as \texttt{nwb4edu}~\citep{voytek_teaching_2020}.


\section{Building our NWB-based data conversion pipeline: Experiences, Challenges, and Lessons Learned}

\subsection{How to organize data into a standard format}

There have been many efforts at standardization of neuroscience data. Neurodata Without Borders (NWB) started as a pilot project to standardize neurophysiology data~\citep{teeters_neurodata_2015}, which then matured into NWB:N version 2.0 (NWB:N 2.0)~\citep{rubel_nwbn_2019}.

However, NWB is not really a file format. The substantive outcome of the NWB development effort was an ``ontology'' that encapsulates the logical structure of neuroscience data at a high level, and schemas to translate these conceptual objects into precise computational objects. Unlike saving an image in JPEG or a document in PDF, to use NWB researchers must make a number of choices specific to their data, with both technical and conceptual implications. 

\autoref{fig:data-standard} illustrates questions faced by researchers who may record multi-modal data scattered across different files and formats. The resulting data need to be organized, unified, and aligned in order to support analysis and collaboration. There can be different strategies to standardize this data, for example from a data lineage standpoint (the choice of the NWB team, \autoref{fig:data-standard}, middle) or from a categorical standpoint (\autoref{fig:data-standard}, right).

Our files mostly follow the default NWB internal structure for optical physiology, though we made our own extension to handle odor data (see \nameref{sec:ndx-odor-metadata}), and argue researchers could benefit from alternative structures, perhaps using aliases or tags, that allow them to interact with their data files following categorical or other organization (see \nameref{sec:wishlist}).

\begin{figure}
\begin{fullwidth}
  \centering
  \includegraphics[trim=1cm 26.5cm 15.5cm 1.5cm, width=0.9\linewidth]{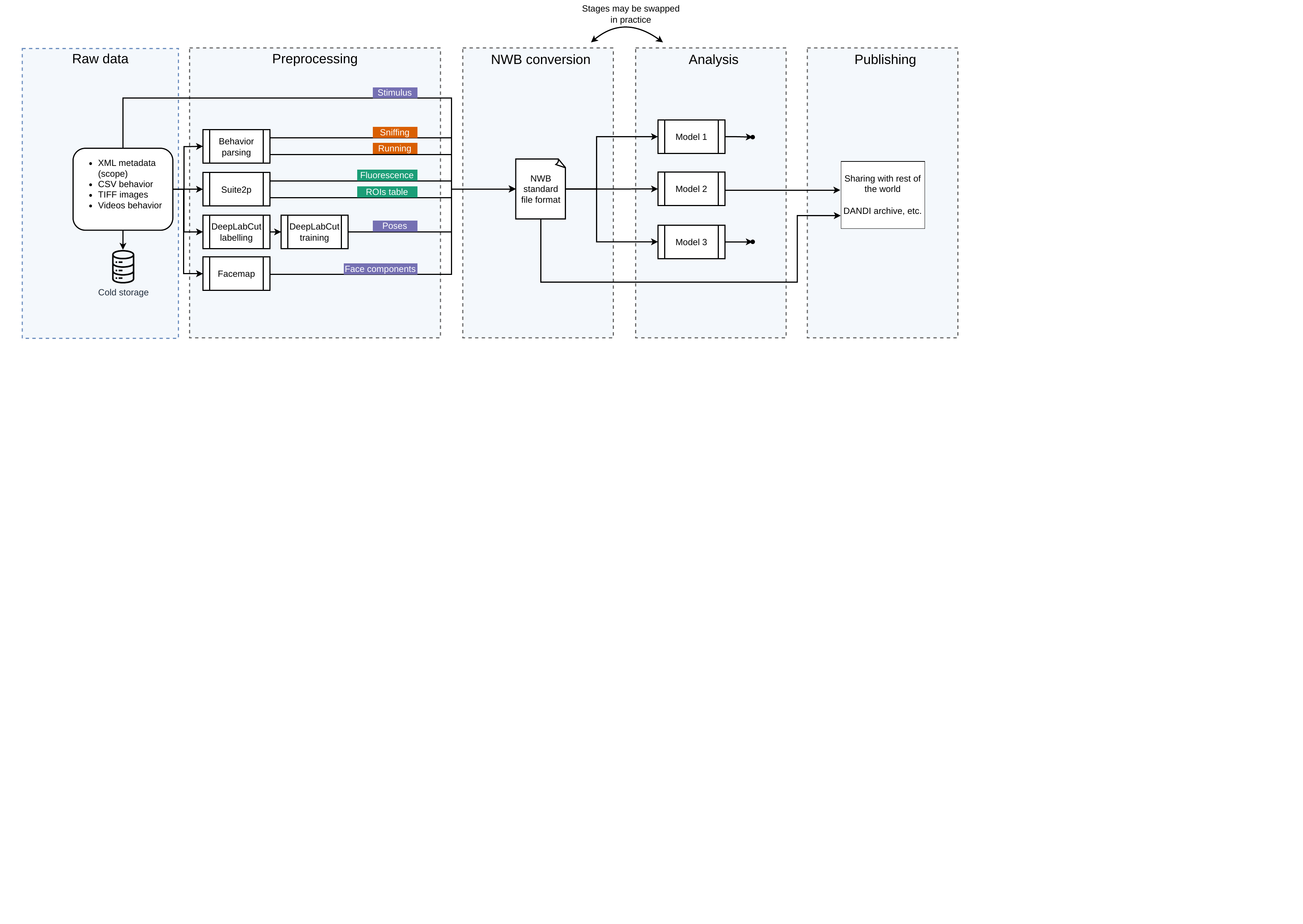}
  \captionsetup{singlelinecheck=off}
  \caption[Our data pipeline]{%
  \textbf{Our data pipeline.} There are five primary stages in our data pipeline. \textit{Raw data} acquired during experiments is archived in cold storage, and also fed to a \textit{preprocessing} stage to be transformed into more directly usable information (\eg{}, fluorescence time series after cell segmentation). This stage uses a range of processing packages that produce multiple files, that are then combined during \textit{NWB conversion} into a standardized format. Scientific \textit{analysis} ideally is performed on the standardized data, but in practice may instead use individual files produced during preprocessing, in which case conversion and analysis stages are swapped. Standardized data is \textit{published}, \eg{} by uploading to a publicly accessible archive, in parallel with traditional journal publication. 
  }%
  \label{fig:data-pipeline}
\end{fullwidth}
\end{figure}


\subsection{When to create and use the standardized format}

Few experimental acquisition systems produce NWB files natively, so use of the standard requires researchers to choose a process and time to convert to NWB from some mixture of other data files. One strategy is to convert at the end of a project, perhaps to upload to a repository for sharing. This choice minimizes disruption to existing research workflows and preserves flexibility for intermediate analyses. However, this strategy may reduce reproducibility, as analysis is done on different files than are eventually shared. Also, shared code needs to be refactored at time of publication to account for these file differences.

Alternatively, conversion could occur prior to internal use. In the pipeline illustrated in \autoref{fig:data-pipeline}, conversion happens between preprocessing (using Suite2p and DeepLabCut) and analysis. Regardless of standardization, researchers typically reformat data before analysis, for example to compile information from multiple raw files into a convenient single data array or table. The key cost of standardization is to place \emph{restrictions} on allowable output formats, in order to reap the benefit of harmonizing a particular dataset with common practice in the field. If data is converted early, then archival repositories can be used also as backups, possibly including data version control. Moreover, shared code does not need substantial rewriting at time of publication. However, if there is not already a robust conversion pipeline in place, this strategy introduces additional effort prior to progress of the scientific aims.

Overall, our feeling is that the stages where NWB is most useful are integrating relatively stable pre-processed data, and archiving finalized data and analysis for publication.

\subsection{Our experience with metadata capture}%
\label{sec:researcher:metadata}

Metadata can be defined as \say{data about data}, for example, information about animal subjects (\eg{} weight, sex, genetic line, age, whether naive or trained), recording sessions  (\eg{} date, task type, experimenter name, manufacturer and model of hardware), stimuli (\eg{} chemical names, concentrations, frequency of audio tones), supplemental text descriptions, and/or parameters used in data processing. Generally, metadata can aid in quality control, communicate contextual information to future users, and support cross-analyses of multiple data sets. Its use can extend beyond the lifetime of a project, including archiving, sharing, and re-use.

\subsubsection{Quality of metadata capture}

A benefit of moving data to NWB is that it encourages systematic handling of metadata. To convert into NWB format, some types of metadata are required by the standard, while some are encouraged. Before moving to NWB, our metadata was scattered in several places. Now, all the relevant metadata is included in the NWB file, allowing consistent and easy access. This may help answer questions such as \emph{What was the sex of animal X?}, \emph{What imaging frame rate was used in experiment Y?}, or, when using our neurodata extension described in \nameref{sec:ndx-odor-metadata}, \emph{Which odor stimulus was used in trial Z?}, without having to go back to the raw data or to the experiment notebook.

\subsubsection{Challenges to metadata capture}

\urldef\SuitepNWBfunction\url{https://github.com/MouseLand/suite2p/blob/118901ac15c6881502c65e011a46fbca16e7a52d/suite2p/io/nwb.py#L346C27-L346C27}

An obvious challenge to incorporating correct metadata in standardized files is that experimentalists do not always record metadata effectively. They may rapidly iterate an experimental design while piloting, and record only ``core'' data for preliminary analyses, with a fuzzy boundary between these initial pilots and subsequent ``real'' data collection. Moreover, metadata often takes unusual effort to document. Acquisition software may not support metadata capture at all. For example, mouse dates of birth or ages are often not included in data files produced during an experiment, yet at least one of these values is needed to create NWB files that meet minimal upload requirements on DANDI (see~\nameref{sec:dandi-surprises}). Sometimes tools set incorrect metadata as a default; for example, we found the NWB conversion function within Suite2p defaulted to setting area of recording to be ``V1''\footnote{\SuitepNWBfunction}.
Also, there is not always a clear purpose to recording metadata that goes beyond the key variables in the original study design. Under the time pressure of the experiment, researchers may be induced either to use non-informative defaults or to enter random metadata to get underway.

This issue is exacerbated by a lack of accepted community standards of how to document for some types of metadata. For instance, in olfaction research, there is not yet consensus on how to document odor stimuli (though see~\citep{castro_pyrfume_2022}, and \nameref{sec:ndx-odor-metadata}).

More generally, metadata capture is needed not only during acquisition but also during preprocessing, analysis, and file conversion stages. Here again a lack of community consensus both motivates the need for detailed metadata capture and illustrates challenges in its implementation. For example, fluorescence is typically normalized, but there is wide variation in how that normalization is performed. Methods used to obtain so-called $dF/F_0$ can differ in parameter choices or the algorithm itself (\eg{} global z-scoring, quantile normalization, or running normalization with additional filtering). Some methods may attempt to compute $dF/\text{noise}$ instead (\eg{} Inscopix CNMFE~\citep{boivin_inscopix_2021}). Often these choices are not apparent in publications and require careful inspection of code, if provided. Such nuances may affect how the data are used, the assumptions of tools that analyze such data, and efforts to replicate analyses.

\subsubsection{Working with acquisition devices and software}

In our labs, research software engineers assist data conversion in part by working with researchers, equipment vendors, and others to determine what metadata is needed and how best to capture it. 

Some commercial vendors put metadata in dedicated files (\eg{}, Bruker Microscope XML or ENV files) while others integrate metadata into the same files as core data (\eg{}, Inscopix Miniscope). However, some proprietary vendor files are poorly documented (and questions stayed unresolved after contacting support), such that we had to reverse-engineer files and make educated guesses as to the information in them. For example, some things we had to independently infer from Bruker XML files were where frame rates are recorded, what physical units different fields have, and what the reference frame coordinates are. Our inferences relied on field names, and were incomplete and possibly in error. More importantly, certain metadata can change the algorithm used to parse a file; for example, a flag indicating whether an experiment has multi-plane imaging affects the correct way to extract timestamps from the XML file. NeuroConv, the conversion tool from NWB developers (see \nameref{sec:neuroconv}), initially did not integrate Bruker metadata~\citep{baker_neuroconv_2023, weigl_add_2023}, but we note support has been added during revision of this manuscript.

Open source tools typically fill a space between commercial vendors and in-lab custom development. Some of these tools lack an ability to input metadata. An example is ArControl~\citep{chen_arcontrol_2017}, which is an experiment control platform used with general purpose microcontrollers to present stimuli and record behaviors. There is a project to convert its output into NWB format~\citep{chen_arcontrol-convert2-nwb_2023}, but (as of this writing) still requiring \textit{post hoc} metadata injection~\citep{chen_issue_2023}.

We also develop custom scripts ourselves that generate CSV-like files on microcontrollers. This approach would ideally include informative headers, for example to give each data column an informative name, a plain text description, physical units, a data type, and possibly other metadata. We find this step introduces friction and an increased chance of errors, especially as experimental designs change and researchers or software engineers need to keep code updated and documented. For now, metadata is often documented after acquisition. In an alternative approach, we implemented custom widgets in Jupyter notebooks used for data acquisition, that allow experimenters to write in odor names.
The notebook then saves the names in a YAML file along with separate core data files, and all files are integrated into an NWB file in a later conversion process. The widget was tedious to develop, but substantially improved the quality of metadata capture for odors at the time of the experiment.

\subsection{Where should raw data and supplemental information be stored?}%
\label{sec:raw-data}

Researchers may want to store raw data in their NWB dataset. In our case, the raw data may contain calcium imaging TIFF stacks or behavior video recordings, both of which tend to be large. For example, a typical calcium imaging session in our lab generates a video of size around 40~GB, with associated behavioral videos around 3~GB. There has long been a question of what to do with videos~\citep{alkan_need_2023, baker_faq_2023}, contrasted with the much smaller data derived from them in pre-processing. Should raw videos be included in NWB files? If yes, how? If not, how should videos be handled when publishing to a repository~\citep{halchenko_dandidandi-cli_2022}?

The NWB team discourages writing videos in lossy compressed formats within NWB files. The main reason is an inability to decode the video without first copying the data to a standard file type (\eg{} MP4) on the user's computer; moreover, if the appropriate codec is not available, even a copied video would be unreadable. The preferred solution is to include videos in NWB files as an \texttt{ImageSeries} that has an external file reference (a relative path to, say, an MP4 file), see~\citep{rodgers_detailed_2022} as an example. This solution also allows adding videos in published datasets on DANDI~\citep{sharda_external_2022}.

Often researchers may want to share explanatory content such as videos of experimental setups or down-sampled videos of calcium imaging registration aligned to behavior recording. Only a subset of recording sessions may have such associated content. A solution could be similar storing raw data as external file references as described above, clearly labelled for demonstrative purposes to avoid confusion.

%

\subsection{How should different data types be stored?}%
\label{sec:custom-data}

In NWB, neurodata types refer to different modalities of data and metadata, for example \texttt{DfOverF}, \texttt{PupilTracking}, or \texttt{SpikeEventSeries}. Each type has specific rules to fit different use cases. If data belongs to a standard neurodata type, there are usually clear examples and guidelines about where and how to store it in an NWB file. When it does not, non-trivial choices may be required, and variation across labs, each implementing their own conventions, may impact general reusability.

For each data source to be integrated into an NWB file, users must answer a number of questions about the data representation. Can the data be fit in a standard neurodata type? What metadata should be associated with it? Would an extension (see \nameref{sec-NWB-extensions}) add a more appropriate datatype? Does such extension exist? If not, is it worth the effort to develop one? 

Additional questions concern where to place the data in the NWB  hierarchy. The organization of the NWB standard is structured with data workflow stages at the top of the hierarchy: \texttt{acquisition} (usually raw),  \texttt{processing}, and \texttt{analysis} (see \autoref{fig:data-standard}). While in theory preserving some element of data lineage, the semantics in practice are not always clear or observed, and can cause confusion when creating and using NWB files.

For example, should raw behavior time series acquired from microcontrollers be in \texttt{acquisition}, a module called \texttt{behavior} in \texttt{acquisition}, or in the same \texttt{behavior} module in \texttt{processing} that is often used to store post-experiment processing such as DeepLabCut pose estimation? From a data lineage point of view, it should be stored in \texttt{acquisition}. But from an analysis point of view, doing so spreads multiple fragments of behavior-related data across multiple hierarchical levels and modules.

\subsubsection{Cell type tagging} As a detailed example of how small experimental variations can lead to non-trivial design choices in NWB files, we describe an experiment in the Fleischmann lab involving two color imaging of red (tdTomato) labelled cells in parallel with green (GCaMP) functional imaging. After using Suite2p for cell segmentation, the researcher classified each cell as expressing or not expressing the red fluorophore, producing a table of ROI (cell) indices, boolean values for whether a cell is red, and auxiliary data about the classification (average pixel intensity and a quality metric).

There are three levels of detail one might choose to keep in an NWB file (in addition to the functional imaging contained in a standard datatype): as the full table, as only the boolean array, or as an array of indices of red cells. The last choice is the most compact, but does not preserve the auxillary information that might be useful for quality control and reproducibility. Similarly, parameters of the classifier itself (\eg{} intensity thresholds) should likely be saved as well. The choice of what information to retain both suggests and is constrained by what datatypes are available, or whether we would need to develop an extension (see \nameref{sec-NWB-extensions}). And a further decision is where to save the data in the file hierarcy (\autoref{fig:data-standard}): as pre-processed data or an analysis result?

There is obvious value to saving the classification in the same place that stored the segmentation table from Suite2p output, essentially by adding more columns to that table. However, since the classification is not available at the time of Suite2p segmentation, and updating existing objects in the Suite2p NWB file was problematic~(see \nameref{sec:issue:composability}), we resorted to placing the classification table in another module called \texttt{cell\_tag}. Given that the table came from Suite2p, whose outputs are in \texttt{processing}, we were unsure whether \texttt{cell\_tag} should be considered \texttt{processing} or \texttt{analysis} in terms of lineage. However, in terms of usage, the tagging is not a useful result by itself, but is combined with the calcium dependent activity. Hence, we decided to consider the table as processed data needed for analysis, and save it in \texttt{processing}.

\subsubsection{Breathing} As a second example, the Datta lab records breathing signals with a temperature sensor implanted in the nose. An Arduino captures the signal, which is written into a CSV file in real-time. We developed a processing pipeline to clean and parse the breathing signal into individual breaths, and store the resulting data in an NWB file. There were a number of challenges along the way that highlight some limitations of the current NWB implementation.

Scipy’s \texttt{signal.find\_peaks} function was the core of the breath processing pipeline; good results relied on choosing correct parameters to find true breaths while ignoring noise in the data. Sometimes we would update the defaults of those parameters based on new analyses, and it would have been helpful to traverse old files programmatically and update them. As it was, many key parameters ended up stored in the \texttt{description} property of the relevant \texttt{TimeSeries}, which may not be an obvious location to those looking at the data for the first time.

Also, there were a number of options for how to store information about each breath, which were difficult to differentiate ahead of time. It would have been ideal to choose based on, \eg{}, efficiency of storage or common practice, but in the end our decision was purely pragmatic. We first considered a tabular format like the \texttt{TimeIntervals} table, but adding data to the \texttt{TimeIntervals} table proved to be cumbersome~\citep{pearl_pandas_2022}. Then we considered an \texttt{IntervalSeries}, which would allow labeling onsets and offsets of inhales and exhales and convey the \say{interval} aspect of the data, but this did not lend itself to storing scalar descriptors for each breath, since the datatype stores only timestamps and not values. Finally, we settled on a simple solution: a \texttt{BehavioralTimeSeries}, containing many \texttt{TimeSeries} of length \texttt{number\_of\_breaths}. For example, inhale onset times, amplitudes, and peak flow rates each got their own \texttt{TimeSeries}. Inhales and exhales were paired in the pre-processing stage, and the \texttt{TimeSeries} that describe the inhales and exhales have the same length, thus implicitly pairing each inhale/exhale pair. We chose to save the \texttt{BehavioralTimeSeries} interface, called \say{breaths}, in the \texttt{processing} section of the NWB file.

\subsection{Should one standardize data from intermediate analysis stages?}

Research analysis pipelines typically have multiple stages, such as pre-processing, statistical modeling, simulation, or any computation whose inputs are the outputs of a previous stage. Those stages may also branch out to test a family of models, or vary analysis parameters. The NWB standard is limited in its handling of analysis parameters, for example as tables of metadata. Should intermediate results be appended to a single NWB file containing the entire history of analysis, each as their own \say{data source}? Should each analysis be stored in its own NWB file? Should all but the final published analysis be discarded? 

Iterative analyses quickly become unwieldy without automated tracking of workflows (\eg{}, Renku\footnote{\url{https://renku.readthedocs.io/}}) and/or data versions (\eg{}, DataLad~\citep{halchenko_datalad_2021}). NWB was not designed to compactly represent collections of results such as arise from parameter sweeping in an analysis. Similarly, NWB does not natively support tracking the partitioning of data, such as into \say{training} and \say{testing} subsets for cross validation (though there are possible \textit{ad hoc} solutions under the current standard, and new packages in late development (personal communication, NWB Developer Team), to support such functionality).

\subsection{Editing and merging of NWB files}%
\label{sec:issue:composability}

Early in our transition to NWB adoption, we needed to combine an NWB file exported from Suite2p with another NWB file produced by our own data pipeline. This turned out to be surprisingly difficult. Indeed, according to the PyNWB documentation, adding to files is supported, but removal and modifying of existing data is not allowed. We therefore tried two approaches to do this. In the first, we read the existing NWB file produced by Suite2p, added the missing data, and exported to a new NWB file. In the second, we looped over containers, \ie{} HDF5 groups, in the existing NWB file, and copied each of them into a new NWB file, together with the new data.

The first approach produced an NWB file that, due to a bug in the underlying packages (which has since been fixed), caused crashes while reading with PyNWB~\citep{pierre_recursionerror_2020}. Because of a different bug, the second approach failed to create a new NWB file with the new containers~\citep{pierre_unable_2020}. These unexpected errors in what seemed like intuitive workflows were frustrating both for the delay in switching over to NWB, and the additional effort needed to diagnose the bugs and find workarounds.

There are still limitations in copying containers from one NWB file to another. But compared to when we started working on this project, it is now more straightforward to copy datasets, \ie{} a data array and its timestamps, from one file to another, and to read an existing NWB file, modify it, and export the modified file to a new file.
It is also possible to append data to a file, in the sense of creating new datasets.
However, to our knowledge, the only way to update metadata in an NWB file is to read the content of the existing file, use the NWB API to create an object with the correct metadata,
and then export to a new file.
In general, we have found that editing and merging NWB files can be a large source of confusion for users, and having a good tutorial or documentation as proposed in~\citep{dichter_documentation_2023} would be extremely useful.

\begin{figure}
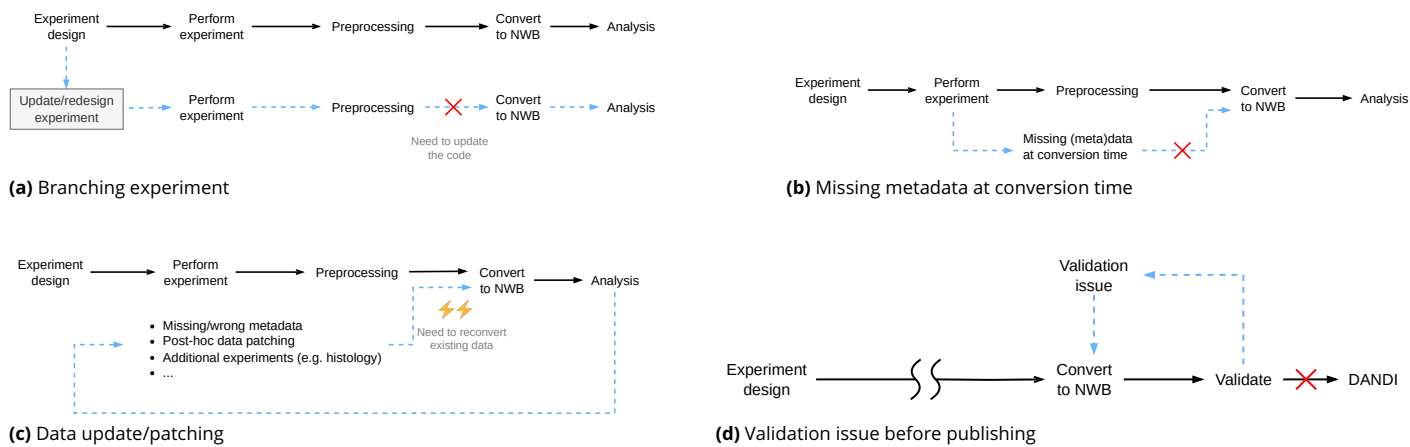

\begin{fullwidth}

    \centering
    \begin{subfigure}[b]{0.45\linewidth}
        \centering
        \includegraphics[page=2, width=\linewidth, trim=0cm 10cm 1cm 0cm]{Figures/workflow-issues-branching.drawio.pdf}
        \caption{Branching experiment}%
        \label{fig:workflow-issues-branching-a}
    \end{subfigure}
    \hfill
    \begin{subfigure}[b]{0.45\linewidth}
        \centering
        \includegraphics[page=3, width=\linewidth, trim=0cm 12cm 1cm 0cm]{Figures/workflow-issues-branching.drawio.pdf}
        \caption{Missing metadata at conversion time}%
        \label{fig:workflow-issues-branching-b}
    \end{subfigure}
    \\[2em]

    \begin{subfigure}[b]{0.45\linewidth}
        \centering
        \includegraphics[page=4, width=\linewidth, trim=0cm 10cm 1cm 0cm]{Figures/workflow-issues-branching.drawio.pdf}
        \caption{Data update/patching}%
        \label{fig:workflow-issues-branching-c}
    \end{subfigure}
    \hfill
    \begin{subfigure}[b]{0.5\linewidth}
        \centering
        \includegraphics[page=5, width=\linewidth, trim=0cm 11cm 3cm 0cm]{Figures/workflow-issues-branching.drawio.pdf}
        \caption{Validation issue before publishing}%
        \label{fig:workflow-issues-branching-d}
    \end{subfigure}
    \\[1em]

    \caption[Pain points scenarios in the conversion workflow]{%
        \textbf{Pain points scenarios in the conversion workflow.} This figure describes different scenarios adding burden to the research workflow. The red crosses represent a situation that breaks the existing workflow. The electric current symbol represents the location of a pain point. \autoref{fig:workflow-issues-branching-a} shows that branching from the main experiment, \ie{} a redesign or update of the experiment, may break the current conversion code to NWB. \autoref{fig:workflow-issues-branching-b} shows that if some metadata is missing at conversion time, it may force the researcher to come back to the experiment, to the original data, or to the conversion code. \autoref{fig:workflow-issues-branching-c} shows a scenario where existing NWB files need to be updated, \eg{} when data from additional experiments like histology experiments become available, or when the NWB files have missing/wrong metadata, or if the NWB file has been found to have some data issues which need to be updated. \autoref{fig:workflow-issues-branching-d} shows a validation issue before publishing the data to DANDI which may force the researcher to update their conversion code to NWB and reprocess their NWB files.
    }%
    \label{fig:workflow-issues-branching}
\end{fullwidth}
\end{figure}


\subsection{Pain points in the conversion workflow}

We encountered several pain points in our data conversion pipeline. One of the main pain points happens with branching experimental designs (\autoref{fig:workflow-issues-branching-a}). Each time a design is updated, NWB conversion code may break and need to be updated. This is an issue especially early in project development, when many experimental details are undecided, but can continue far into a project's lifetime as researchers adjust their approach based on prior results.

Another pain point may arise when metadata is missing at conversion time (\autoref{fig:workflow-issues-branching-b}). Researchers may be tempted to input nonsense values that need to be updated later, or the conversion may be blocked until the missing metadata is captured.

Sometimes, data in NWB files may need to be updated, \eg{} to correct a previous entry, or  to add data that becomes available later, such as histology (\autoref{fig:workflow-issues-branching-c}). In this case, the pain point happens when the data conversion pipeline has to be run again on multiple already existing files. As discussed more in \nameref{sec:dandi-validation-surprises}, a related issue can arise when sharing data in an archive such as DANDI~\citep{halchenko_dandidandi-cli_2022}. Validation to DANDI is stricter than requirements to build a file with the python API (PyNWB), requiring conversion code updates even after conversion was locally \say{successful} (\autoref{fig:workflow-issues-branching-d}).

\subsection{Timeliness of code contribution acceptance}

\begin{figure}
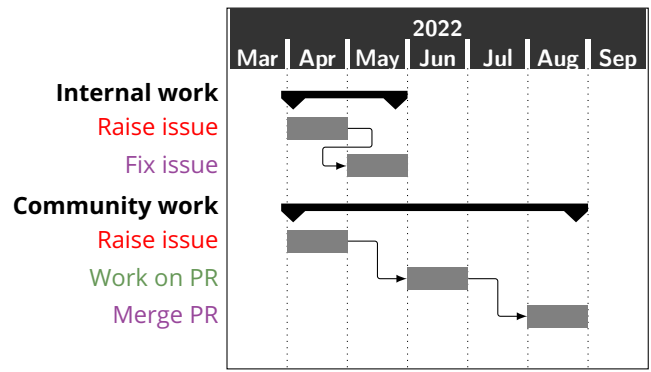

    \centering
    \begin{adjustbox}{max width=\columnwidth, keepaspectratio}
    \begin{ganttchart}[
        y unit title=0.4cm,
        y unit chart=0.5cm,
        x unit=0.8cm,
        vgrid,
        time slot format=isodate-yearmonth,
        time slot unit=month,
        title/.append style={draw=none, fill=white!20!black},
        title label font=\sffamily\bfseries\color{white},
        title label node/.append style={below=-1.6ex},
        title left shift=.05,
        title right shift=-.05,
        title height=1,
        bar/.append style={draw=none, fill=white!50!black},
        bar height=.6,
        bar label font=\normalsize\color{black!50},
        group right shift=0,
        group top shift=.6,
        group height=.2,
        group peaks height=.3,
        bar incomplete/.append style={fill=Maroon}
        ]{2022-03}{2022-09}
        \gantttitlecalendar{year, month=shortname} \\
        \ganttgroup{Internal work}{2022-04}{2022-05} \\
        \ganttbar[name=T1A]{\textcolor{red}{Raise issue}}{2022-04}{2022-04} \\
        \ganttlinkedbar{\textcolor{Purple}{Fix issue}}{2022-05}{2022-05} \\

        \ganttgroup{Community work}{2022-04}{2022-08} \\
        \ganttbar[name=T2A]{\textcolor{red}{Raise issue}}{2022-04}{2022-04} \\
        \ganttlinkedbar{\textcolor{OliveGreen!75}{Work on PR}}{2022-06}{2022-06} \\
        \ganttlinkedbar{\textcolor{Purple}{Merge PR}}{2022-08}{2022-08} \\
    \end{ganttchart}
    \end{adjustbox}
    \caption{\textbf{Example of a broader community issue resolution timeline.} This figure illustrates the time taken to fix a Suite2p-related issue internally (\ie{} two months), compared to the time it took to fix the issue for the broader community (\ie{} five months).}%
    \label{fig:chan2-issue-timeline}
\end{figure}

We discovered Suite2p was dropping data from a second microscope channel in its NWB file output. The issue was that the NWB export function had been developed for only one microscope channel. \autoref{fig:chan2-issue-timeline} shows the timeline of the issue until a fix was released. While fixing the issue internally took around two months, it took around five months (including time for us to complete a GitHub \say{pull request}) for the solution to be available to the Suite2p community. This is a long turnaround for what we considered to be a critical error, impacting all multicolor imaging analysis. We stress that we appreciate the Suite2p team's review and acceptance of our code contribution. However, this experience illustrates a general problem for research software development in the open source community; researchers maintaining software may not have the bandwidth to address every issue or feature request in as timely a fashion as desired.

\subsection{Off the shelf NWB conversion}%
\label{sec:neuroconv}

Some friction during adoption of NWB can arise from the level of technical skill needed to be able to convert one's data. When we started the process of adopting NWB, the options available were either to learn how to write our own data conversion pipeline, or hire a consultant to do the technical work. In the few years since, the NWB ecosystem has rapidly evolved. More recently introduced tools miss some areas of need (\eg{} currently unsupported proprietary formats like Inscopix, or Suite2p output with multiple channels) , but they solve many popular use cases.

NeuroConv~\citep{baker_neuroconv_2023} is a rapidly advancing Python package from core NWB developers to make it easier to convert from a variety of common neuroscience data formats. It is a flexible low-code solution for use in one-off conversion or as part of a lab pipeline. One benefit of NeuroConv is that it includes utilities to get metadata from proprietary formats with minimal effort. Additionally, it can combine files from multiple data sources with functionality to align timestamps, and contains utilities for file path inference to aid batch-conversion based on user-defined data organization. Coupled with the development of the NWB Graphical User Interface for Data Entry (NWB GUIDE)~\citep{m_nwb_2023}, which uses NeuroConv as a back-end, NWB is considerably more accessible to newcomers than it was at the time we began our adoption.

These recent changes highlight a risk to early adopters of any standard, that one may build features from scratch that quickly become obsolete after further developments from the community. If we started this project today, we would leverage these community projects, developing less custom code and using existing features from more widely tested projects used by the entire NWB community.

\subsection{An indirect benefit of using NWB is improved data awareness}

As a standard, NWB encourages good data practice. For example, each data array that is written in a file needs to have a timestamps vector attached to it. And ideally all the timestamps of the same NWB file would be on a common axis, which can be quite challenging for experiments with multi-modal recording and has been discussed recently in~\citep{rodgers_detailed_2022}.
This includes the acquisition timezone, meaning an NWB file can easily be analyzed in different parts of the world without risking timestamps collision.

In our case, standardization encouraged better timestamping with custom instruments and sensors like Arduino and Teensy boards. For example, before we developed our own data pipeline, one lab researcher manually specified inter-trial intervals in their analysis code, as it was cumbersome to extract the (nearly constant) intervals from the recording system. Now they have access to the actual recorded timestamps for the inter-trial intervals and can catch and correct any system errors. Also, using NWB encouraged us to align timestamps across all data sources, simplifying downstream analysis work.

A general by-product of moving our lab to NWB is increased awareness regarding data management itself. Lab members have become more familiar with general principles such as FAIR~\citep{wilkinson_fair_2016} and emerging best practices. Although still harboring some skepticism of the direct usefulness to their research, lab members have become more welcoming to incorporating NWB into their workflows, and are supportive of the broader benefits, such as for data sharing.

\section{Creating NWB extensions allows fitting domain specific use cases}%
\label{sec-NWB-extensions}

An emerging standard with as broad a domain as NWB will naturally struggle to cover some applications, especially in less common experimental settings. Making the standard extensible creates a way for individual users or research groups to add functionality beyond what is created by the core developers. The NWB standard thus includes \say{neurodata extensions} to incorporate new data types. Extensions may be used individually, shared with the community, or, if the extension addresses a fundamental gap in NWB coverage, submitted for review to be added to the standard NWB data types. We have had some success using and creating NWB extensions to fit our specific research needs, though challenges and questions remain.
\keypoints{motivation and definition of extensions}

\subsection{Existing Neurodata Extensions}

Before deciding to create an extension, researchers should check the Neurodata Extensions Catalog (NDX Catalog), a community led effort to create a central repository of contributions that, by design, arise from widely distributed effort~\citep{ruebel_ndx_2023}. The NDX catalog includes extensions that support diverse types of data such as TTL pulses~\citep{ly_ndx-events_2023}, and popular acquisition systems such as miniscopes~\citep{dichter_ndx-miniscope_2023}. However, not all Neurodata Extensions are listed on the NDX Catalog, since anyone can create and post an extension on lab websites, GitHub, or other sites.

\subsection{Lab-specific metadata}

One use case of NWB extensions is to record lab-specific metadata with greater flexibility than is supported in base NWB. We created \texttt{ndx-fleischmann-labmetadata}~\citep{pham_ndx-fleischmann-labmetadata_2023} to store additional detail on recorded brain areas, and descriptions of the experiment and animals. Within our general type of experiment we use many variations (\nameref{box:lab_workflow}), such as 1-photon or 2-photon calcium imaging, single or multicolor imaging, head-fixed or freely-moving animals, and passively presented or task-driven stimulation. NWB standard is missing fields to describe some of the complexity in these experiments; for example, we use multicolor imaging to retrograde label projections from the imaging site to distant brain regions, and there is no field to indicate this second (projection) area. Storing such additional experimental description as text in the top-level description field would be harder for quality control at time of entry, and less efficient to parse for queries at analysis time. With our extension, a subset of information ends up being repeated with standard locations in the NWB file; for example, imaging site is also stored under \texttt{ophys}, as suggested in the NWB documentation. However, we chose to centralize our metadata in one place, to make querying, analysis, and aggregation of multiple data files easier.
\keypoints{lab metadata list + benefits}

\subsection{Odor stimulus metadata}%
\label{sec:ndx-odor-metadata}

Another use case for extensions is to describe stimuli that do not fit within base NWB types. Our calcium imaging experiments use primarily odor stimuli, and some non-chemical stimuli such as sound. We are not aware of an extension to adequately describe these stimuli, and hence a year ago developed \texttt{ndx-odor-metadata}~\citep{pham_ndx-odor-metadata_2023}.  We characterize odor stimulus with standardized information automatically obtained from PubChem~\citep{kim_pubchem_2023} using a PubChem CID (chemical IUPAC names, molecular formulas, and weights); dilution details such as concentration and solvent; metadata that are useful for analysis such as stimulus category (\eg{} control or conditioned stimulus) and common chemical names; and identifiers to cross-reference with associated time series. The extension also allows non-odor stimuli to be described in plain text.
\keypoints{odor metadata extension details of what's included, and in relation to data analysis}

A major challenge with such extension development, although not an issue specific to NWB, is that there may not be community consensus or documentation to be used as starting points for extension design. For odor stimuli, it was not obvious what type and level of description would be necessary for both in-lab analysis and general reproducibility. Fleischmann lab RSEs used existing spreadsheets as starting examples, and learned only later that outside collaborators had independently created a package, \texttt{pyrfume}~\citep{castro_pyrfume_2022}, for documentation of odorants. Future work could better harmonize these two efforts at stimulus metadata capture. More generally, the technical development of metadata capture can grow only in concert with the research community's understanding of what the standards for metadata ought to be.
\keypoints{challenge: lack of community standards as starting points, solution: look at examples}

\subsection{Documentation for extension development}%
\label{sec:ndx:challenge-during}

For most labs, we expect extension development will be out of reach unless the lab has access to personnel with strong coding experience. A general challenge for us was that the available documentation could be confusing, and information was scattered across multiple sources, including  documentation pages for PyNWB\footnote{\url{https://pynwb.readthedocs.io/}},
HDMF\footnote{\url{https://hdmf.readthedocs.io/}}~\citep{tritt_hdmf_2019}, NWB Overview\footnote{\url{https://nwb-overview.readthedocs.io/}},
and NWB Schema\footnote{\url{https://nwb-schema.readthedocs.io/}},
and also in GitHub issues or examples on Slack. It would have been helpful in particular to have a larger set of use cases, examples, and/or tutorials. We stress that the NWB development team was highly responsive through GitHub, Slack, and emails, and their help was very valuable for our development work. In the future, we hope such support could be complemented by more comprehensive documentation.

\subsection{Social challenges in extension development}%
\label{sec:ndx:challenge-after}

One lesson learned from our experience is that creating the extension is only a technical part of a solution. Sustained engagement with researchers to choose, document, and record key information is the more fundamental requirement, especially if metadata standards motivating the extension are unsettled. 

As a lab, we continue to refine what metadata we should track and how we should capture it. Some changes arise from variation in experiments conducted by different lab members. Some changes reflect interest in adding further types of information, such as water restriction details for experiments with behavioral training, as inspired by an International Brain Lab extension~\citep{baker_ndx-ibl_2023}. An extension may lower the technical barrier to metadata capture, but only if the extension is aligned with researchers' goals and practices, including changes over time.

A closely related challenge is that many metadata records must be captured \textit{post hoc} instead of automatically during acquisition or pre-processing. Some acquisition systems lack features to enter metadata in machine-readable formats (necessary for software to correctly place that information in NWB files) during the experiments. Even where real time capture is possible, the systems may be cumbersome to use, leading researchers to avoid comprehensive entry and checking of metadata. We usually need to work with researchers to collect metadata records in machine-readable formats after experiments and preprocessing are completed, leading to increased work and greater risk of errors and missing information.

We also have felt a tension between building minimal extensions that serve immediate needs versus investing in a longer development project that may have greater generalizability.  For example, our odor stimulus extension provides for single odorant but not mixed odor stimuli. Though we generally do not use multi-component odors, they are used by some of our close collaborators~\citep{wilson_primacy_2017}. We also designed our extension to build on PubChem standardization, which presents difficulties when studying custom-made or undocumented natural odors~\citep{li_musk_2022}. These limitations in our current implementation may become impediments as neuroscience tends towards more natural and ethologically relevant behaviors~\citep{krakauer_neuroscience_2017}. However, surmounting these challenges will require substantial engagement from a broad section of the olfaction research community, before any technical contributions such as extensions can have a substantial impact.

\subsection{Framework extensions}

An extension is built on top of another NWB object. This object can be one of the four minimally structured objects (Groups, Attributes, Links, Datasets) of the base NWB specification~\citep{ruebel_nwb_2020}, but it is often better for an extension to build on a previously developed high level data type that already captures much of the structure of the information being added. In addition to making it easier to develop the extension without starting from scratch, such inheritance can promote greater consistency by keeping almost all data organization the same as a ``common'' data type, except for the particular items added by the new extension. For example, a new fluorescence imaging data type might add beam path parameters to an existing fluorescence imaging type, to provide for a scope that uses non-uniform laser scanning but otherwise collects standard data.

In cases where NWB is missing a more basic category of data, there is motivation to develop extensions intended to be used specifically as building blocks for other extensions.  We refer to these types of building blocks as ``framework extensions''. In addition to facilitating development and serving as illustrative examples, framework extensions could add technical precision to discussions if a research community is working to converge to a consensus data standard.

For example, DeepLabCut and Facemap output time series of spatial locations of points on an animal's body. While these outputs can be stored generically as simply behavior, they are both instances of a more specific concept of ``pose'', and can be stored using the \texttt{ndx-pose} extension~\citep{ly_ndx-pose_2022} (the DLC developers offer the \texttt{DLC2NWB} utility to ease conversion using this extension, but we are not aware of an analogous tool for Facemap).

An example framework extension that could have broad utility would store results from principal component analysis (PCA) (one of the authors, TP, participated in discussing this idea at a 2023 NWB Hackathon, but it is not yet implemented as far as we know). PCA is used widely as a simple data dimensionality reduction technique. There are several variants of PCA, such as jPCA used to find low dimensional structure in the activity of large neural ensembles~\citep{churchland_neural_2012}. Moreover, many analysis applications, including Facemap and MoSeq~\citep{wiltschko_mapping_2015, wiltschko_revealing_2020, lin_dattalabmoseq2-app_2023}, use PCA as a preprocessing step. A general PCA extension could serve as a useful framework to incorporate these different uses within a consistent NWB format. The framework extension would define component eigenvalues, eigenvectors, and projections of the original time series.


As another example, BEADL\footnote{\url{https://beadl.org/}}
and ArControl~\citep{chen_arcontrol_2017} model behaviors in a finite state machine framework. The extension \texttt{ndx-beadl}~\citep{ly_ndx-beadl_2021} is available for BEADL outputs, and it is possible to adapt the extension to handle ArControl output~\citep{chen_arcontrol-convert2-nwb_2023}. However, as finite state machines are an important class of models for analysis, there could be value in establishing a more general framework extension, for example called \texttt{ndx-finite-state}, from which extensions for these specific analysis packages would inherit.

\subsection{Wishlist for NWB extensions}

Development, cataloging~\citep{ruebel_ndx_2023}
and updating extensions could be more streamlined.

\begin{figure}
\begin{fullwidth}
    \centering
    \includegraphics[width=\linewidth]{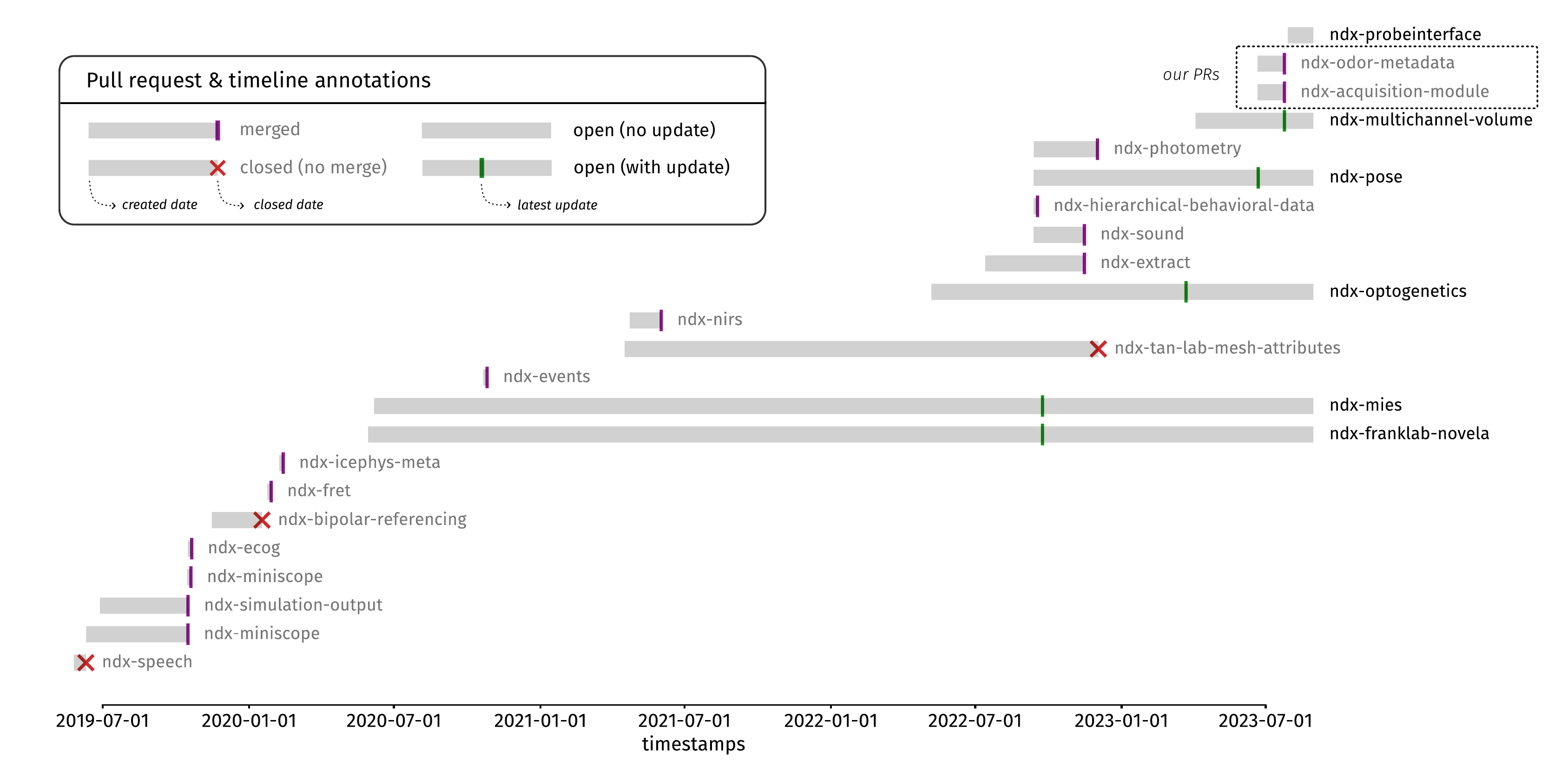}
    \caption[Pull requests (PR) for publishing on the extension catalog may take a long time to be accepted]{%
        \textbf{Pull requests (PR) for publishing on the extension catalog may take a long time to be accepted}.
        The data were obtained using GitHub API from
        \href{https://github.com/nwb-extensions/staged-extensions}{\texttt{nwb-extensions/staged-extensions}}
        repository, on \DTMdate{2023-07-30}.
        Out of $23$ extension requests, about $61\%$ ($14/23$) have been merged (bars ended with purple vertical sticks) and added to the catalog, while $13\%$ ($3/23$) are closed without being added to the catalog (bars ended with red crosses).
        The review times for finished PRs vary, ranging between within a day to less than $5$~months for most of them, with the exception being $1.6$~years for the closed request for \texttt{ndx-tan-lab-mesh-attributes}.  
        About $26\%$ of the extension PRs ($6/23$) are still open, with $3$ out of $6$ being stale for more than a year. 
        A notable one is \texttt{ndx-pose} for pose estimation extension (PR $\#31$) which has been open for almost a year (Sept. 2022).
        Note: any closed/merged PR finished within less than $5$ days is artificially extended to be $5$ days for visibility.
    }%
    \label{fig:staged-PR}
\end{fullwidth}
\end{figure}


First, researchers may develop software using different repository hosting (\eg{} GitLab instead of GitHub). It could be more inclusive for the \texttt{ndx-template} extension template~\citep{ly_ndx-template_2019} to not explicitly assume GitHub as the code repository. The template might also take into account both the Python Package Index (PyPI) and Anaconda as potential package repositories.

Second, currently, to be added to the NDX Catalog, new extensions are submitted via Pull Requests for review on GitHub. Some seem to be approved instantly while others are either stale (\eg{} \texttt{ndx-pose}), or took around 2 months to be approved (see \autoref{fig:staged-PR}). While the timeline for open source development is often highly variable, researchers and RSEs have to balance many priorities, and usually cannot dedicate much time to the approval process.

To simplify the review process, a bot could check critical requirements before asking for intervention from an NWB maintainer (taking some inspiration from the Conda-Forge community). For example, the bot could check if the package is already published on PyPI, if all the metadata fields in the \texttt{ndx-meta.yaml} file are filled in, and if all tests pass. Also, the bot could help for updating the extensions, say if the extension template or if some dependency has changed. Also, publishing to PyPI could be streamlined, for example by having a CI job in the \texttt{ndx-template} extension template~\citep{ly_ndx-template_2019} that supports automatic publishing to PyPI.

Additionally, we suggest adding some metadata to improve quality checks, centralization, and organization of extensions. To maintain quality control, the catalog could allow entries to be tagged to indicate whether an extension has been reviewed, similar to the distinction of pre-print from peer-reviewed publications. To tackle fragmentation of extensions and tools, it might be helpful to also allow optional specification of the type and lineage of each entry, \eg{} whether it is built upon another extension, and if it is a template extension for demonstration purposes. Additionally, we found it unclear whether the catalog submission policy welcomed lab-specific extensions (\eg{} \texttt{ndx-ibl} for the International Brain Laboratory (IBL) and ours \texttt{ndx-fleischmann-lab}), though in comments on an earlier draft, the NWB team clarified that they do encourage such submissions (personal communication). Although lab specific, these extensions could be useful examples or starting points for other labs to develop their own.

We hope to see depositing on the community catalog become more flexible and timely. A disadvantage is a potential reduction of quality control. However, more engagement, contribution, feedback, and discussion from the community is in general more likely to accelerate development of the standard. Extensions may serve as a starting point for such discussions, responding to community needs.

\section{Considerations for sharing on DANDI}%
\label{sec:dandi-surprises}

In this section, we look at the last step of the data conversion workflow: data has already been converted to NWB and the researcher wants to share the data on a public respoitory, for example to accompany a published paper. Here we look at DANDI~\citep{halchenko_dandidandi-cli_2022}, as the default solution recommended by the NWB team.

\subsection{Potential surprises with data validation}%
\label{sec:dandi-validation-surprises}

One possible source of friction is validating the data before being able to push to DANDI. DANDI enforces a set of rules that NWB files have to meet before upload and publication as a \say{dandiset} is allowed, intended to promote adherence to consistent metadata standards and ensure the FAIRness~\citep{wilkinson_fair_2016} of the archive. If files do not meet those requirements, researchers may need to (iteratively) redo their conversion with altered settings. This can be an unpleasant surprise, as one might have thought that having converted to NWB itself would be sufficient.

One solution could be to promote and describe the NWBInspector tool~\citep{baker_nwb_2020}, used to validate NWB files, in the documentation and tutorials on how to create NWB files. It would also be helpful to be able to run NWBInspector from PyNWB to check files and get feedback at the time of initial conversion. This solution may soon be implemented when using no-code tools like NWB-GUIDE~\citep{m_nwb_2023} (see also \nameref{sec:neuroconv}), though it did not exist when we started our projects.

Another point of friction can arise if a dandiset has already been published but needs to be updated later, for example~\citep{pierre_update_2023}. In our case, the validation rules changed after we first released the dandiset, and files that were already published became retroactively non-compliant. We had to go back to conversion from raw data. In general, if the cost to update a dandiset is too high, the risk is that researchers may decide not to correct stale or inaccurate information.

\begin{figure}
    \centering
    \includegraphics[width=0.9\linewidth]{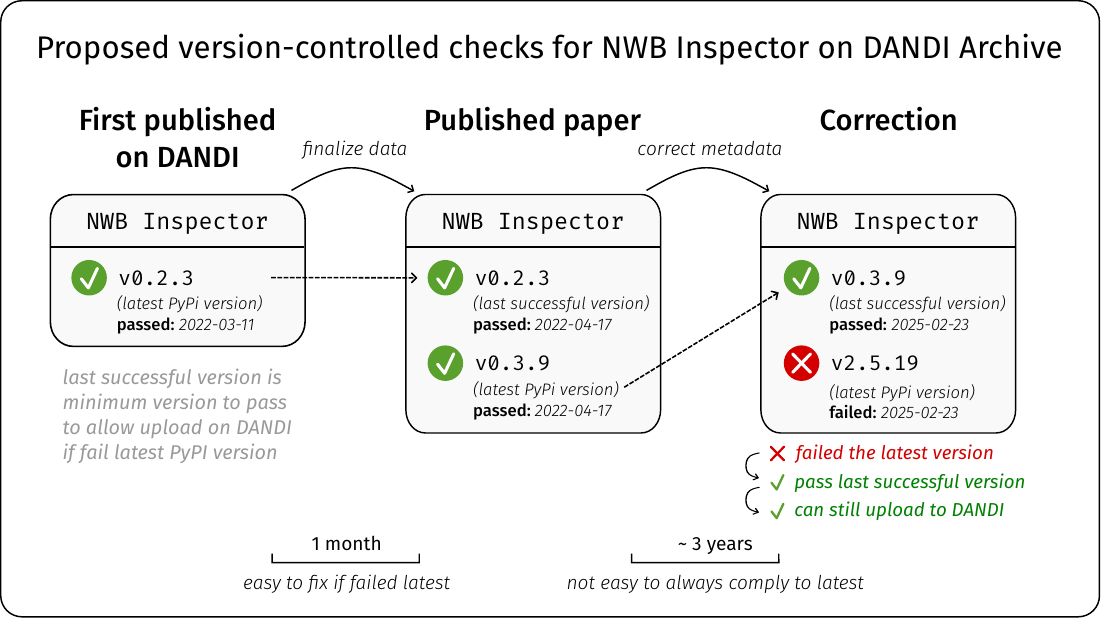}
    \caption[Proposed version-controlled checks for NWBInspector when uploading to DANDI Archive]{%
        \textbf{Proposed version-controlled checks for NWBInspector when uploading to DANDI Archive}.
        To be published on DANDI Archive, datasets should always be checked and pass the latest version of NWB Inspector (first and second boxes), to maintain compliance with best practices.
        When existing datasets need to be updated, they may fail the latest version, for example 3 years after publication, to correct metadata (third box on left).
        The proposed solution is to allow for checking against the \emph{last working version} for existing datasets, in cases of non-compliance with the latest version. 
        This solution still allows researchers to disseminate updates and corrections, while maintaining transparency for the community in terms of non-compliance.
        This solution can be allowed a limited number of times, and failures can also be reported to DANDI Archive maintainers.%
    }
    \label{fig:dandi-inspector}
\end{figure}

A potential solution would be to allow version-controlled inspection (\autoref{fig:dandi-inspector}). There could be at least two levels of NWBInspector passing. Files that pass the most recent NWBInspector can always be uploaded. But if some files already on DANDI get updated and fail the most recent inspection, they could still be uploadable given they passed the previous working version of NWBInspector. Similar to CI systems, logs of fail/pass versions could be attached to the archive for developers and others to inspect. This approach would allow for researchers to flexibly upload corrections and updates, while still being transparent about compliance status. Failures could be reported to the DANDI team, allowing them to work with researchers to follow up-to-date best practices.

\subsection{Modification of file organization}

Another potential surprise is that the DANDI upload tool renames and reorganizes files into a \say{flatter} hierarchy. For example, one could have their NWB files organized by experiments with a nested directory structure organized by areas of recording, but DANDI refactors this structure to be organized only by subject directories, and moreover renames files by subject name and data type. DANDI also modifies external file links stored inside each NWB file to stay consistent with these file changes.

Changing the file structure may break existing analysis pipelines based on the original paths. Thus, it may be useful to think about data archiving from the start of a project. In that case, publishing the data to DANDI from the beginning of the project, with occasional updates, would make the researcher aware of this reorganization and account for it in their own code. In addition to saving effort at publication time, such a workflow would enhance analysis reproducibility. However, the cost is some increased overhead while data collection is still occurring.

\subsection{Alternatives to DANDI and general strategy with data repositories}

DANDI has strong restrictions on data file formats. While there is currently an exception on DANDI \citep{rodgers_detailed_2022} that contains free-form source data (\eg{} Python and NPY files), it is unclear whether this feature will officially be supported in the long run. Alternative repositories include Zenodo\footnote{\url{https://zenodo.org}},
Figshare\footnote{\url{https://figshare.com/}},
GIN G-Node\footnote{\url{https://gin.g-node.org/}},
OSF~\citep{foster_open_2017}, or university data storage, potentially with Globus endpoints~\citep{foster_globus_1997, foster_globus_1998, foster_globus_2006}. An alternative decentralized solution is Academic Torrents~\citep{cohen_academic_2014, lo_academic_2016}, which uses the BitTorrent protocol and leverages university bandwidth to avoid unsustainable data storage costs over the long term. These data archives can include NWB data and all related data such as raw data, pre-conversion data, analysis and summary data.

However, it may not always be feasible to centralize all data, and researchers might instead use a multi-site storage strategy. Large source data, including raw and pre-conversion data, could be deposited on university storage solutions, with Globus endpoints if possible, to take advantage of universities' generally less restrictive quotas, assuming these data would rarely be accessed, updated, or used after conversion. Converted NWB files could then be deposited on DANDI, on which researchers can benefit from specialized software tools, as well as DANDI Hub, a Jupyter Hub with free computing resources on Amazon Web Services (AWS). Lastly, along with code and documentation, researchers could continuously work on data with their analysis pipelines using solutions such as GIN G-Node, GitHub/GitLab with a DataLad~\citep{halchenko_datalad_2021} or DVC~\citep{skshetry_dvc_2023, barrak_co-evolution_2021} backend, to manage aggregated and analyzed data and code. This helps with version-controlled code and data, without the restrictions from DANDI Archive.

We note that if researchers decide to follow a multi-site strategy, they would need to manually link these different archives together, preferably with DOI numbers and in machine-readable metadata on these different providers. The outlined example strategy separates the three archives (\eg{} university storage, DANDI Archive, and GIN G-Node) by an assumed increasing update frequency, \ie{} raw data files are less frequently updated compared to NWB files, and NWB files less than files with analysis or modelling results. With distributed storage, especially if these assumptions do not apply, researchers would need to manually keep track and link the updates regularly.

\section{Suggestions to streamline data reading and writing}%
\label{sec:user-exp}

\subsection{Data exploration tool guidance}

The NWB ecosystem has many applications available for a researcher to quickly get a sense of what is inside an NWB file. As of writing, there are four general and 15 specialized data tools listed on the NWB Overview\footnote{\url{https://nwb-overview.readthedocs.io/}},
and new tools continue to emerge. The number of active projects indicates a vibrant development community. However, new users may be overwhelmed by the choices, and not know how, except through brute force trials, to determine which tools are best for them. Moreover, consolidation around a few key applications could help channel valuable developer efforts into refining and improving existing tools, some of which still exhibit rough spots like freezing on large files or frequent crashes.

This situation is common in open source development ecosystems (for example, there are many partially redundant but not interchangeable python plotting packages). A difference here is that the NWB standard was created and continues to be maintained through a somewhat centralized development team, with an explicit agenda to be adopted as a ubiquitous standard for neurophysiology. There is thus a stronger case that innovation arising from widely dispersed development should be balanced by centralized advising over third party tools.

For example, primary NWB documentation could maintain a section with some (automatically scraped) metrics for each tool (\eg{} number of GitHub stars, number of downloads on PyPI) next to accessible summaries of the features of each tool, and descriptions of who their target users are. At time of writing, several of these changes are in process or planned (NWB Team, personal communication)\footnote{\url{https://nwb-overview.readthedocs.io/en/latest/tools/analysis_tools_home.html}}.

A more assertive approach would select recommended tools, on the basis of features, robustness (\eg{} resolution of bugs, handling of large file sizes), and probable longevity. For data exploration, some natural candidates could be NWBWidgets~\citep{dichter_neurodatawithoutbordersnwbwidgets_2019}, which is also integrated with DANDI Hub, and relatively newer NeuroSift~\citep{dichter_neurosift_2023}, which is an interactive visualization tool that works directly in the user's browser. In our experience, NeuroSift is highly accessible, without requiring installation, and offers strong visualization functionality out of the box. Both tools support streaming data from the DANDI Archive. Again, the goal would be to provide soft incentives that encourage contributors to focus primarily on existing tool refinement, while still leaving space for new specialized projects in early development.

\subsection{Data access pain points}

\subsubsection{Figuring out where data is}%
\label{sec:how-to-extract-data}

We find new NWB users often struggle to find and access information, with confusion arising from where the information is in the internal hierarchy, or because the datatype of a particular object does not intuitively describe what it is. Many scientists look first for modules based on source of data (\eg{} fluorescence, behavior, stimuli). But access under the NWB schema runs first through stage of processing (\eg{} acquisition, pre-processing, analysis) and then descends through multiple levels of hierarchy to data source. That is, researchers may employ a mental sequence of \emph{where is my behavior} (say) then \emph{what processing has been applied}, which is the opposite ordering from what NWB currently uses (\autoref{fig:data-standard}).

An outlier is that \texttt{stimulus} is at the top of the hierarchy, with \texttt{acquisition} and \texttt{processing}. However, stimulus time series sometimes need additional processing, for example, to transform raw digital outputs recorded by behavior control devices into a semantically useful tabular format. Should such stimuli be saved within \texttt{stimulus} (with processing stage indicated in \texttt{name} or \texttt{description} attributes), or in a module inside \texttt{processing}? Additionally, tables cannot be saved inside \texttt{stimulus}, and only limited metadata can be associated. It is recommended to use dedicated modules or objects designed to save metadata, for example \texttt{devices} for recording or \texttt{lab\_metadata} for lab-specific  metadata. This again runs into the potential issue of categorically similar objects being widely separated.

    









\begin{figure}
\begin{fullwidth}
    \noindent\begin{minipage}[t]{.47\linewidth}
        \begin{lstlisting}[
            caption=Retrieving data using the PyNWB API,
            label=lst:nwb-api,
            language=Python,
            ]
# 1D array of timestamps
t = nwb_file.processing["behavior"]["interpd_500"]["therm_highpassed"].timestamps[:]

# 1D array of data
therm = nwb_file.processing["behavior"]["interpd_500"]["therm_highpassed"].data[:]
        \end{lstlisting}
    \end{minipage}\hfill
    \begin{minipage}[t]{.47\linewidth}
        \begin{lstlisting}[
        caption=Retrieving data through a custom wrapper,
        language=Python,
        label=lst:wrapper,
        ]
# Use wrapper to create an alias to the data
interpd = MyCustomWrapper(
    nwb_field="processing",
    path_to_interface=["behavior", "interpd_500"],
    nwb_file=nwb_file
)

# After one-time setup, simpler data retrieval
t, therm = interpd.get("therm_highpassed")
        \end{lstlisting}
    \end{minipage}
    \caption[Code snippet comparison]{Code snippet comparison showing how to retrieve data from an NWB file using the \say{raw} PyNWB API (\autoref{lst:nwb-api}) compared to using a custom wrapper (\autoref{lst:wrapper}). After a one-time setup, retrieving the data through a custom wrapper reduces the cognitive load for the user.}%
    \label{fig:code-wrapper}
\end{fullwidth}
\end{figure}

\begin{figure}
\begin{fullwidth}
    \centering
    \includegraphics[width=0.8\linewidth]{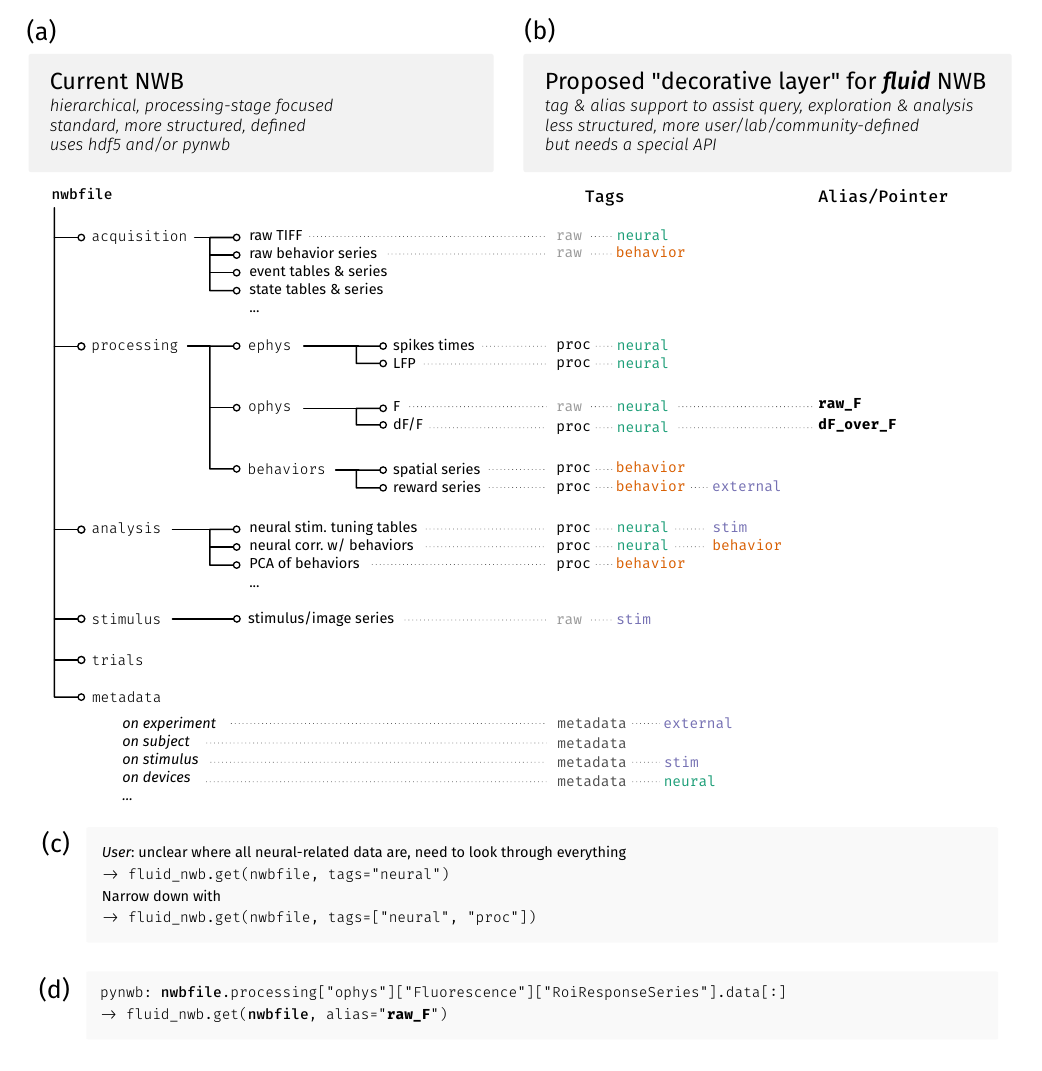}
    \caption[A proposed design layer for the NWB standard to assist with data retrieval and organization]{%
        \textbf{A proposed design layer for the NWB standard to assist with data retrieval and organization}.
        The nature of the current NWB structure is hierarchical and tends to be organized by processing stages; panel (\textbf{a}) shows an example of this structure.
        Accessing relevant data requires knowledge of where it is located, which may be multiple levels deep, see for example bottom box (\textbf{d}) to access raw fluorescence data with \texttt{PyNWB}.
        The proposed ``decorative layer'' allows for more ``fluid'' interaction with NWB via additional specifications in NWB objects, to assist querying, exploration and analysis with more user/lab/community's control and customization, without breaking the existing hierarchical NWB structure.
        Panel (\textbf{b}) illustrates examples of adding tags and aliases.
        Tags can be more specific, multi-faceted and customized to concepts of recording/analysis 
        that users tend to look for (\eg{} \emph{neural}, \emph{behavior}, \emph{stim}, \emph{external}), as well as higher level details such as processing stages (\eg{} \emph{raw}, \emph{proc}).
        Aliases and/or pointers allow users to add names for objects that are most frequently accessed, or expected to be so.
        Taking advantage of this ``decorative layer'', users and developers may design a \texttt{fluid\_nwb} API to interact with NWB files in a more flexible and less verbose manner,
        for example with tags in box (\textbf{c}) and aliases in box (\textbf{d}).
        }%
    \label{fig:fluid-nwb}
\end{fullwidth}
\end{figure}

%

\subsubsection{Cumbersome syntax to extract data}

A challenge for new users that is parallel to understanding object locations is confusion over the addressing syntax, \ie{} when to use dot syntax, \texttt{object1.object2}, or Python dictionary syntax, \texttt{object1[\-"object2"]}. The syntactic variation derives from the structure of the HDF5 file specification and the NWB schema, both of which are generally unknown and opaque to users.

Two obvious alternative possibilities for API syntax would simply make one or the other access method universal (\eg{}~through a Python DataClass). Either choice would obscure the real differences between types of objects in the NWB implementation (\eg{} a fluorescence object including metadata attributes, vs a numpy array just of the $dF/F_0$ values), but we are not convinced that most users benefit from having these differences encoded in syntax.

Another possibility that is both general and convenient for programmatic access would support universal reference via \say{path strings}, such as \texttt{nwbfile[pathstr]} where \texttt{pathstr='object1/\-object2/\-object3'}.

\subsubsection{Lab specific wrapper workaround}

In its current state, long hierarchies in NWB files (\eg{} processing $\rightarrow$ behavior $\rightarrow$ interpolated $\rightarrow$ position $\rightarrow$ data) are slow to type and hard to remember, and tend to clutter code. A common method to hide complexity in an individual user's analysis code is to first create \say{wrappers} (\autoref{fig:code-wrapper}). For example, a wrapper may define simple \say{\texttt{get()}} methods that automatically skip parts of the object path, \eg{} \texttt{data=nwb\_wrapper.get(''dFF0'')}. Wrappers can also add convenience features, such as aggregating different time series into a single data frame, and, wrappers can be stored in dictionaries for easy looping over multiple files.

On the other hand, wrappers may be complex to design and may introduce a maintenance burden if they aim to work across the usually wide range of experiments and data streams that arise even within a single lab. In practice, then, individual researchers often end up partially or completely rewriting similar helper code with each new project.

\subsubsection{Suggestion for better data access: tags and aliases}%
\label{sec:wishlist}

A potential solution for better data access is a feature we call \say{fluid NWB} (\autoref{fig:fluid-nwb}), allowing for a list of tags for each object, including \say{flat} objects such as timeseries, tables, and modules. Users could add annotations and categories as they see fit, and specialized communities could evolve their own norms for \say{virtual} file organization, without confounding the underlying standard. Aliases, to our knowledge, are currently not possible, but the integration of such a feature may allow for users to have easier and quicker access, and could also aid documentation. For example, the AllenSDK has a dedicated dictionary for metadata field mapping to NWB/HDF5 locations; this shares some similarity with aliasing and illustrates a place for annotation usage\footnote{\url{https://alleninstitute.github.io/AllenSDK/allensdk.core.brain_observatory_nwb_data_set.html\#allensdk.core.brain_observatory_nwb_data_set.BrainObservatoryNwbDataSet.FILE_METADATA_MAPPING} (accessed on 2023/12/07)}. Supporting custom tags for neurodata types is currently an open GitHub issue~\citep{ly_support_2022}.

Tags and aliases would be a \say{decorative layer} on top of the NWB standard, allowing for more \say{fluid} data structures, which researchers and developers could exploit for usability and discoverability. However, in the absence of convergence on naming norms within a given research area, overlapping tags, complex tag formatting, and tag relations could proliferate to the point of no longer being useful. For example, should cardiac recordings (EKG), saccades, and arena locations all carry a common \textit{behavior} tag? Should muscle recordings (EMG) be tagged both as \textit{neural} and \textit{behavior} in a brain-machine-interface (BMI) study? The added flexibility of an alias or tag system would produce the greatest benefit if complemented by a process to secure community consensus around tagging conventions.

\section{Conclusion}

Standardization is an essential component of modern data management, analysis, and sharing, and NWB has introduced a comprehensive and versatile data science ecosystem for neuroscience research. However, our experience suggests that implementation of NWB workflows at the level of individual labs or research collaborations still requires significant effort and commitment. Furthermore, given the rapid pace of technology development in neuroscience research, we expect that the development and implementation of adequate data science tools will continue to pose new challenges for some time. Solutions to these challenges will likely require a reorganization of neuroscience research to facilitate interdisciplinary collaborations, including additional institutional support not just for the creation of new tools, but also for their adoption by research labs at all levels of technical capability.

\section*{Acknowledgments}%
\addcontentsline{toc}{section}{Acknowledgements}%
\label{sec:acknowledgements}
We would like to thank Simon Daste and Max Seppo for their input and provision of experimental data. We thank the Osmonauts (U19NS112953), Cindy Poo, Chris Rodgers, Rebecca Tripp, and Emilya Ventriglia for helpful comments on earlier drafts. We thank the NWB, DANDI, and CatalystNeuro teams, including Cody Baker, Ben Dichter, Garrett Flynn, Satrajit Ghosh, Yaroslav Halchenko, and Oliver R\"{u}bel, for extensive discussions and helpful comments following posting of a preprint draft on arXiv. Work in the Fleischmann and Datta labs was supported by NIH award U19NS112953. Work in the AF lab was also supported by NIH award R01DC017437, and the Robert J and Nancy D Carney Institute for Brain Science. Carney Institute computational resources used in this work were supported by the NIH Office of the Director award S10OD025181.

\FloatBarrier

\bibliography{references.bib}








\end{document}